\PassOptionsToPackage{unicode}{hyperref}
\PassOptionsToPackage{hyphens}{url}
\PassOptionsToPackage{dvipsnames,svgnames,x11names}{xcolor}
\documentclass[
 12pt]{article}

\usepackage{amsmath,amssymb}
\usepackage{placeins}
\usepackage{multirow}
\usepackage{makecell}
\usepackage{dsfont}
\usepackage{iftex}
\usepackage{algpseudocode}
\usepackage{algorithm}
\ifPDFTeX
 \usepackage[T1]{fontenc}
 \usepackage[utf8]{inputenc}
 \usepackage{textcomp} 
\else 
 \usepackage{unicode-math}
 \defaultfontfeatures{Scale=MatchLowercase}
 \defaultfontfeatures[\rmfamily]{Ligatures=TeX,Scale=1}
\fi
\usepackage{lmodern}
\ifPDFTeX\else 
\fi
\IfFileExists{upquote.sty}{\usepackage{upquote}}{}
\IfFileExists{microtype.sty}{
 \usepackage[]{microtype}
 \UseMicrotypeSet[protrusion]{basicmath} 
}{}
\makeatletter
\@ifundefined{KOMAClassName}{
 \IfFileExists{parskip.sty}{%
  \usepackage{parskip}
 }{
  \setlength{\parindent}{0pt}
  \setlength{\parskip}{6pt plus 2pt minus 1pt}}
}{
 \KOMAoptions{parskip=half}}
\makeatother
\usepackage{xcolor}
\makeatletter
\ifx\paragraph\undefined\else
 \let\oldparagraph\paragraph
 \renewcommand{\paragraph}{
  \@ifstar
   \xxxParagraphStar
   \xxxParagraphNoStar
 }
 \newcommand{\xxxParagraphStar}[1]{\oldparagraph*{#1}\mbox{}}
 \newcommand{\xxxParagraphNoStar}[1]{\oldparagraph{#1}\mbox{}}
\fi
\ifx\subparagraph\undefined\else
 \let\oldsubparagraph\subparagraph
 \renewcommand{\subparagraph}{
  \@ifstar
   \xxxSubParagraphStar
   \xxxSubParagraphNoStar
 }
 \newcommand{\xxxSubParagraphStar}[1]{\oldsubparagraph*{#1}\mbox{}}
 \newcommand{\xxxSubParagraphNoStar}[1]{\oldsubparagraph{#1}\mbox{}}
\fi
\makeatother

\usepackage{longtable,booktabs,array}
\usepackage{calc} 
\usepackage{etoolbox}
\makeatletter
\patchcmd\longtable{\par}{\if@noskipsec\mbox{}\fi\par}{}{}
\makeatother
\IfFileExists{footnotehyper.sty}{\usepackage{footnotehyper}}{\usepackage{footnote}}
\makesavenoteenv{longtable}
\usepackage{graphicx}
\makeatletter
\def\maxwidth{\ifdim\Gin@nat@width>\linewidth\linewidth\else\Gin@nat@width\fi}
\def\maxheight{\ifdim\Gin@nat@height>\textheight\textheight\else\Gin@nat@height\fi}
\makeatother
\setkeys{Gin}{width=\maxwidth,height=\maxheight,keepaspectratio}
\makeatletter
\def\fps@figure{htbp}
\makeatother

\makeatletter
\@ifpackageloaded{caption}{}{\usepackage{caption}}
\AtBeginDocument{%
\ifdefined\contentsname
 \renewcommand*\contentsname{Table of contents}
\else
 \newcommand\contentsname{Table of contents}
\fi
\ifdefined\listfigurename
 \renewcommand*\listfigurename{List of Figures}
\else
 \newcommand\listfigurename{List of Figures}
\fi
\ifdefined\listtablename
 \renewcommand*\listtablename{List of Tables}
\else
 \newcommand\listtablename{List of Tables}
\fi
\ifdefined\figurename
 \renewcommand*\figurename{Figure}
\else
 \newcommand\figurename{Figure}
\fi
\ifdefined\tablename
 \renewcommand*\tablename{Table}
\else
 \newcommand\tablename{Table}
\fi
}
\@ifpackageloaded{float}{}{\usepackage{float}}
\floatstyle{ruled}
\@ifundefined{c@chapter}{\newfloat{codelisting}{h}{lop}}{\newfloat{codelisting}{h}{lop}[chapter]}
\floatname{codelisting}{Listing}

\makeatother
\makeatletter
\@ifpackageloaded{caption}{}{\usepackage{caption}}
\@ifpackageloaded{subcaption}{}{\usepackage{subcaption}}
\makeatother

\ifLuaTeX
 \usepackage{selnolig} 
\fi
\usepackage[]{natbib}
\usepackage{bookmark}

\IfFileExists{xurl.sty}{\usepackage{xurl}}{} 
\hypersetup{
 pdftitle={Title},
 pdfauthor={Author 1; Author 2},
 pdfkeywords={3 to 6 keywords, that do not appear in the title},
 colorlinks=true,
 linkcolor={blue},
 filecolor={Maroon},
 citecolor={Blue},
 urlcolor={Blue},
 pdfcreator={LaTeX via pandoc}}

\newcommand{\anon}{1}

\usepackage{xparse}

\NewDocumentCommand{\anonflag}{m}{%
  \ifnum\anon=1\relax
    #1%
  \fi
}

\newcounter{suppsection}
\renewcommand{\thesuppsection}{S\arabic{suppsection}}
\newcommand{\suppsection}[1]{%
  \refstepcounter{suppsection}%
  \section*{\thesuppsection\quad #1}%
}
\newcounter{subsuppsection}[suppsection]
\renewcommand{\thesubsuppsection}{\thesuppsection.\arabic{subsuppsection}}
\newcommand{\subsuppsection}[1]{%
  \refstepcounter{subsuppsection}%
  \subsection*{\thesubsuppsection\quad #1}%
}
\usepackage{caption}
\usepackage{titlesec}
\titlespacing{\section}{0pt}{1.5ex plus 0.5ex minus 0.1ex}{1ex}
\titleformat{name=\paragraph,numberless}[runin]
 {\normalfont\bfseries}{}{0pt}{}
 
\titlespacing*{\paragraph}{0pt}{0.5ex plus 0.2ex minus 0.1ex}{1em}
\begin{document}

\def\spacingset#1{\renewcommand{\baselinestretch}%
{#1}\small\normalsize} \spacingset{1}


\if1\anon
{
 \title{\bf MoSAIC: Multi-Resolution Spatial Regression Analysis of Cellular Colocalizations in Cancer Imaging}
 \author{
  Jessica Aldous$^{1}$ ,
  Michele Peruzzi$^{1}$,
  Maria Masotti$^{2}$,
  Aaron Udager$^{3}$,\\
  Allison May$^{4}$,
  Evan Keller$^{5}$,
  and
  Veerabhadran Baladandayuthapani$^{1}$
  \\
  \small
  $^1$Department of Biostatistics, University of Michigan
  \\
  \small
  $^2$ Hennepin Healthcare Research Institute 
  \\
  \small
  $^3$Department of Pathology, University of Michigan\\
  \small
  $^4$Department of Urology, University of Virginia
  \\
  \small
  $^5$Department of Urology, University of Michigan
}
 \maketitle
} \fi

\if0\anon
{
 \bigskip
 \bigskip
 \bigskip
 \begin{center}
  {\LARGE\bf \bf MoSAIC: Multi-Resolution Spatial Regression Analysis of Cellular Colocalizations in Cancer Imaging}
\end{center}
 \medskip
} \fi

\begin{abstract}

Hierarchical multiplex imaging approaches generate spatially resolved single-cell measurements across multiple, spatially organized fields of view (FOVs) within patient tumor specimens, thereby enabling systematic investigation of how the organization of the tumor microenvironment varies along biologically meaningful intratumoral gradients. Existing approaches fail to jointly address this multi-resolution data structure needed to recover true biological signals. We propose \texttt{MoSAIC}: \textit{multi-resolution spatial regression analysis of cell colocalizations}, a hierarchical Bayesian spatial regression model designed for multi-resolution spatial data. \texttt{MoSAIC} decomposes the joint variation into three model components: (i) global tumor-gradient effects, (ii) patient-specific effects to capture inter-patient variability, and (iii) Gaussian process models to account for spatial dependence between FOVs within each patient tumor tissue. Simulations demonstrate \texttt{MoSAIC} has improved prediction and model fit compared to existing spatial and non-spatial model alternatives. Our method is motivated by and applied to a renal cell carcinoma multiplex imaging cohort to investigate immune–tumor colocalization patterns across the epithelial-to-mesenchymal transition (EMT) gradient. \texttt{MoSAIC} identifies increased macrophage–tumor colocalization and decreased cytotoxic T–tumor colocalization progressing across the increasing EMT gradient, consistent with EMT-associated immune suppression and spatially varying immune engagement. Overall, \texttt{MoSAIC} provides an interpretable, multi-resolution framework for quantifying spatial tumor-gradient effects in cancer imaging studies. Software is available on Github\anonflag{ at \href{https://github.com/jcaldous/MoSAIC}{jcaldous/MoSAIC}}.
 
\end{abstract}

\noindent%
\vfill

\newpage
\spacingset{1.8} 
\section{Introduction}\label{sec-intro}
\begin{figure}[ht!]
 \centering
 \includegraphics[width=1\linewidth]{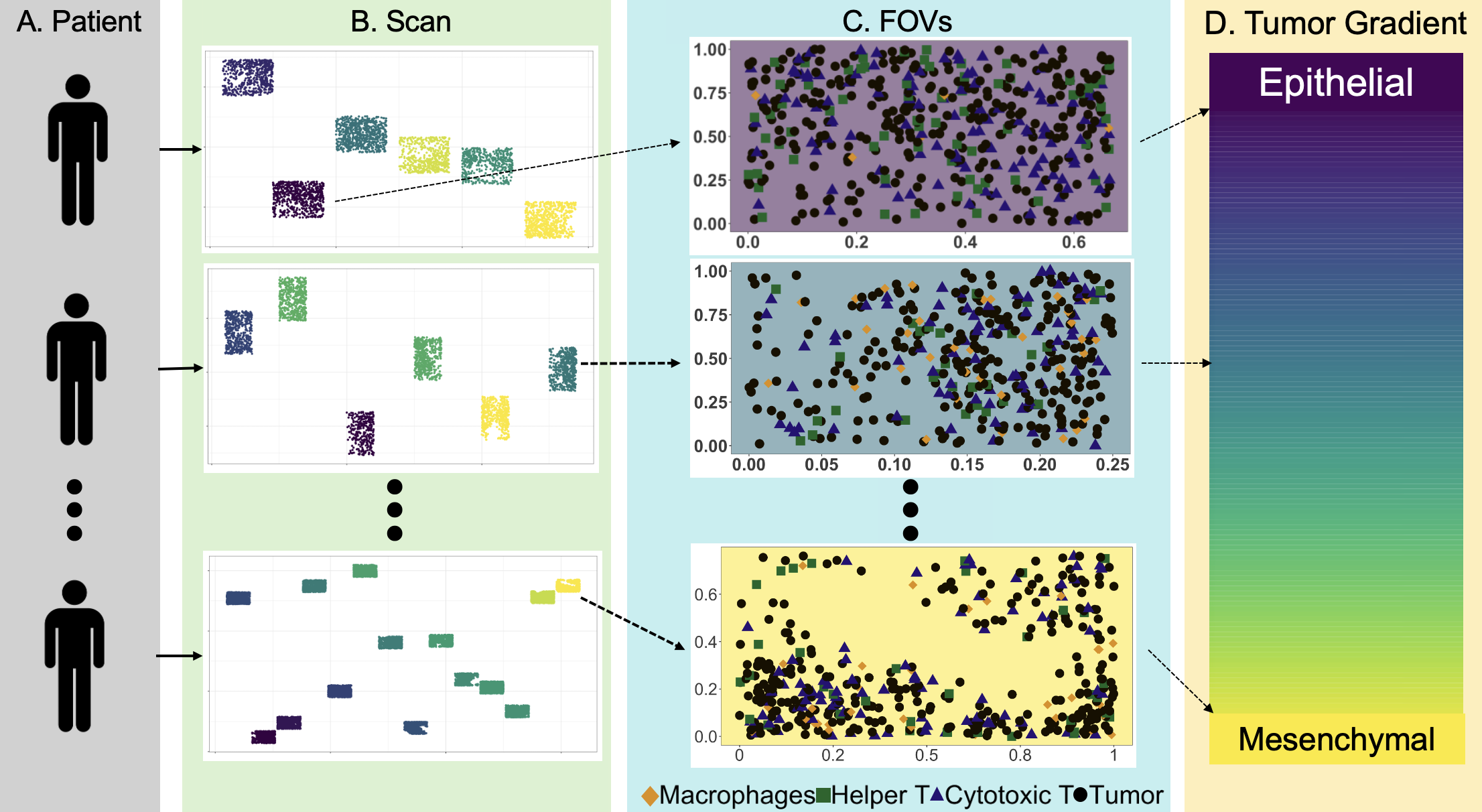}
 \caption{\textbf{Hierarchical multiplex imaging data structure in sRCC}. Each patient (A) has a tumor tissue scan with multiple FOVs (B). Each FOV (C) is composed of 4 different cell types (macrophages: orange diamonds, helper T: green squares, cytotoxic T: navy triangles, Tumor: black circles). FOVs are colored based on a continuous biomarker (D) of the EMT gradient (purple: clear cell; yellow: sarcomatoid).}
 \label{fig:alt_fig_1}
\end{figure}

\paragraph*{Influence of tumor microenvironment in cancer.}
As cancer develops, tumor cells alter their environment to support survival and progression (\citealt{TMEdef}). This interplay between tumor cells, immune cells and the surrounding stroma forms a complex tumor ecosystem, termed the tumor microenvironment (TME; \citealt{Boyk2022Tumor}). Evidence suggests spatial positions of immune cells within the TME can influence drug resistance (\citealt{vandammultiplex2022}) and either promote or suppress tumorigenesis (\citealt{TMEgen}). For example, a high infiltration of M2 macrophages, white blood cells involved in wound healing, can promote tumor growth and is associated with poorer patient prognosis (\citealt{TAM_ref}). In contrast, cytotoxic T cells target tumor cells by identifying abnormal antigens and an abundance of this immune cells is generally associated with better prognosis (\citealt{TMEgen}). Recent developments in spatial multiplex imaging technologies, such as tissue-based cyclic immunofluorescence (\citealt{Linmethod}) and multiplexed ion beam imaging (\citealt{ionbeam}), have enabled a more systematic and quantitative investigation of the TME (\citealt{lewisspatial2021}). These technologies generate rich datasets with measures of phenotypic and functional protein markers at the cellular level while maintaining the spatial relationships between cells within the tumor. 
\paragraph*{Sarcomatoid renal cell carcinoma.}
In the context of kidney cancers, particularly renal cell carcinoma (RCC), studies have delineated the critical role of the TME and its components in RCC biology and progression (\citealt{Heideggerrcc}). Of particular interest is the involvement of the TME in initiating the epithelial to mesenchymal transition (EMT), a biological process that causes epithelial cells to progressively lose their typical features for a mesenchymal-like phenotype, marking significant changes in cell invasiveness (\citealt{emtref}). The EMT is associated with sarcomatoid dedifferentiation in RCC (sRCC), an aggressive form of renal cancer (\citealt{pdl1sarc}). Importantly, sRCC is not a unique histological subtype but rather a cell dedifferentiation that can occur anywhere in the tumor. Thus, one tumor contains cells at various stages along the EMT gradient. Although the EMT is a well-described oncogenic pathway, the role of the microenvironment in initiating this transition to more aggressive disease remains poorly understood in RCC. Recent studies also revealed that sRCC is increasingly sensitive to immunotherapy compared to clear cell RCC, suggesting that immune-tumor cell colocalization may differ across the EMT gradient (\citealt{TAM_ref}). 
 Consequently, our primary scientific question focuses on how cellular colocalizations vary \textit{across} the tumor gradient both within and across patient tumors. 
\paragraph*{Tumor gradients and hierarchical multiplex imaging data.} Spatial multiplex imaging data offers an ideal platform for scientific investigations of the tumor gradient, as it measures both spatial biomarker expression and cellular spatial relationships. Since tumor tissues are large and heterogeneous, studies often focus on sampling smaller subsections of the tumors, called fields of view (FOVs). These spatially resolved replicates enable the study of the TME both within and across patients in a cost-effective manner, resulting in a hierarchical data structure that we call Hierarchical Multiplex Imaging (HMI). The general structure of HMI data from our sRCC data is presented in Figure \ref{fig:alt_fig_1}. Each patient (A) has a tumor tissue scan with spatially organized FOVs sampled across the two-dimensional tumor tissue space (B). Each FOVs has a corresponding spatial distribution of the cells (C), wherein each shape and color represents a unique cell type. FOVs are colored by a continuous biomarker that tracks the EMT across the tumor tissue (D). Our scientific focus is on quantifying changes in cellular organization across this EMT gradient. 

\paragraph*{Existing methods and limitations.}
This HMI data structure presents several unique analytical challenges. Specifically, (a) the complex spatial relationships among cells in a single FOV; (b) the non-ignorable spatial dependency between FOVs of a single tumor tissue; and (c) the high degree of both within- and between-patient variability (heterogeneity), which must all be considered to quantify true biological gradient within and across the patient tumor tissues. Existing methods for addressing some of these challenges can be broadly split into two categories: summary-based and model-based. Several summary-based methods have been proposed (\citealt{Wrobel2023}), and they typically quantify cellular colocalizations within each image (FOV) that can be utilized in downstream cross-patient analyses (\citealt{summarymethods}). Examples include cross functions like Ripley's K-cross (\citealt{Ripley1976}) and G-cross (\citealt{SPATT}), density-based methods like DIMPLE (\citealt{MASOTTI2023100879}), and others like MIAMI (\citealt{MIAMI}) which models the phenotypic expression directly as opposed to the discrete cell types. These summary-based models are well-suited for characterizing immune-tumor cellular colocalizations within a \textit{single} sample/FOV but ignore (spatial) relationships between FOVs. As a result, they are not readily scalable or adaptable to the hierarchical dependence structures of HMI data.

Existing model-based methods are limited in their ability to account for every source of variability. Standard mixed effects regression models (\citealt{Galecki2013}) can borrow strength across patient data, but they ignore spatial dependencies among FOVs, which can lead to artificially low p-values and inflated effect sizes (\citealt{shakedpvalues,lennonpvalues}). Neighborhood-based methods such as conditional autoregressive (CAR) models (\citealt{besag1991bayesian}) are poorly suited for HMI since tumor tissues are non-conformable spaces; patient tumors have different shapes, EMT gradients, and numbers of FOVs leading to very different neighborhoods (\citealt{hierspac}). Gaussian Processes (GP; \citealt{RasmussenWillGP}) offer a more flexible alternative to capture the spatial relationships between FOVs, but currently available GP models for cellular colocalization were not developed to accommodate spatially resolved within-patient replicates, making them difficult to implement (\citealt{SVCA}). Few models (e.g. \citealt{spaveanova}; \citealt{spicyr}; \citealt{eliasonrao2024}; \citealt{bipps}; \citealt{Kondo}) borrow information across patients with multiple images to investigate differences in cell colocalization across patient cohorts. However, none of these methods account for a hierarchical relationship within and between images, and their simplifying assumptions make their direct application unsuitable for HMI data. 

\paragraph*{Multi-resolution spatial regression analysis of cell colocalizations in cancer.}
To this end, we propose \texttt{MoSAIC}: \underline{m}ulti-res\underline{o}lution \underline{s}patial regression \underline{a}nalys\underline{i}s of cell \underline{c}olocalizations. \texttt{MoSAIC} seeks to bridge the gap between these approaches, culminating in a procedure that disentangles the layers of within- and between-patient variability to produce interpretable estimates. Specifically, we employ a hierarchical Bayesian spatial regression model that quantifies the effect of the tumor gradient on the TME while accounting for the spatial relationship between the FOVs and the between-patient heterogeneity. We model the tumor gradient nonlinearly using spline-based basis functions and incorporate GP-based random effects to capture spatial dependencies among FOVs. The resulting model produces interpretable coefficient curves that highlight how cellular colocalizations change across the tumor gradient and appropriately accommodate between-patient variability. In multiple simulation settings, \texttt{MoSAIC} demonstrates improved predictive performance and model fit compared to spatial and non-spatial model alternatives for HMI data. When applied to sRCC HMI data, \texttt{MoSAIC} detects alterations in the spatial colocalization patterns of immune cells—specifically cytotoxic T cells and macrophages—with tumor cells along the EMT gradient.

Our paper is organized as follows. Section \ref{sec-meth} details the \texttt{MoSAIC} model, including model construction, estimation, and posterior inference. Section \ref{sec-sim} outlines the comparative simulation studies, followed by Section \ref{sec-rd} that applies \texttt{MoSAIC} to the sRCC HMI data. Section \ref{sec-conc} ends with some concluding thoughts and a discussion on model extensions and generalizations. Supplemental material covers additional modeling details, simulation results, and figures. Software for fitting \texttt{MoSAIC} is freely available on Github\anonflag{ at \href{https://github.com/jcaldous/MoSAIC}{jcaldous/MoSAIC}}. 

\section{\texttt{MoSAIC} Model}\label{sec-meth}
\paragraph*{HMI data structure and notations.}
We denote each patient by $i=1,\dots,I$ and their tumor tissue space by $ \mathcal{D}_i$. Within each $ \mathcal{D}_i$, we sample a set of $n_i$ non-overlapping FOVs indexed by $j (i)=1,\dots,n_i$. We represent each FOV by their centroid locations, $\boldsymbol{s}_{j (i)}$, which denote the patient subspace $\mathcal{S}_i\subset \mathcal{D}_i$. In our setting, we have two hierarchical spatial resolutions in our sRCC HMI data: (a) a fine-scale resolution consisting of cells, which are nested within (b) a coarser-scale resolution of FOVs sampled across a given tumor tissue for a given patient. We develop \texttt{MoSAIC} to leverage the FOV level information to investigate the influence of the tumor gradient across the entire tumor tissue. Our patient subspace $\mathcal{S}_i$ is comprised of FOVs sampled across the patient tumor tissue $\mathcal{D}_i$ but \texttt{MoSAIC} can be applied to any data set with spatial locations across many replicates (e.g. cells within an FOV).\\
For a given patient subspace $ \mathcal{S}_i$, let $y (\boldsymbol{s}_{j (i)})$ be our outcome measured at $\boldsymbol{s}_{j (i)}$. We denote our $p$ covariates measured at $\boldsymbol{s}_{j (i)}$ as $\boldsymbol{x} (\boldsymbol{s}_{j (i)})=\{x_1 (\boldsymbol{s}_{j (i)}),\dots,x_p (\boldsymbol{s}_{j (i)})\}$. In our RCC data, $y (\boldsymbol{s}_{j (i)})$ is a scalar quantifying the degree of immune-tumor cellular interactions for an FOV with centroid $\boldsymbol{s}_{j (i)}$, and $\boldsymbol{x} (\boldsymbol{s}_{j (i)})$ denotes a vector storing the expression level of p biomarkers within an FOV representative of the tumor gradient. To better estimate complex global (population level) relationships between $\boldsymbol{x} (\boldsymbol{s}_{j (i)})$ and $y (\boldsymbol{s}_{j (i)})$, we borrow strength across the patient cohort while accounting for within patient (spatial) heterogeneity. To this end, we estimate the effect of the $\boldsymbol{x} (\boldsymbol{s}_{j (i)})$ on $y (\boldsymbol{s}_{j (i)})$ as follows:
\begin{gather*}
\begin{aligned}\label{equ_tumor tissue}
 y (\boldsymbol{s}_{j (i)}) &= {g{(\boldsymbol{x}(\boldsymbol{s}_{j (i)}))}}+ \mu_i+ \psi (\boldsymbol{s}_{j (i)})+ \epsilon_{j (i)},\\
 \end{aligned}
 \end{gather*}
where ${g{(\boldsymbol{x}(\boldsymbol{s}_{j (i)}))}}$ is the global covariate effect capturing the tumor gradient, $\mu_i$ is the patient-specific intercept, $\psi (\boldsymbol{s}_{j (i)})$ is the spatial random effect, and $\epsilon_{j (i)}$ captures the residual measurement error. 

Aggregating data across all I patients, let $\boldsymbol{y}_{N\times1}=[ \boldsymbol{y}^T(\boldsymbol{\mathcal{S}}_1),\dots,\boldsymbol{y}^T(\boldsymbol{\mathcal{S}}_I)]^T$ denote the concatenated vector of outcomes. For ease of exposition, we represent the covariates of interest across all $I$ patients by matrix $\boldsymbol{X}_{N\times p}$ and the block-diagonal design matrix that maps the vector of patient-specific intercepts ($\boldsymbol{\mu}_{I\times1})$ to the $N$ total observation by $\boldsymbol{Z}_{N\times I}$. We also vectorize the spatial random effects ($\boldsymbol{\psi}_{N \times 1}$) and measurement errors ($\boldsymbol{\epsilon}_{N \times 1}$). The resulting population level \texttt{MoSAIC} model is as follows:
\begin{gather}\label{equ_pop}
 \begin{aligned}
 \boldsymbol{y}=&\boldsymbol{g}(\boldsymbol{X})+\boldsymbol{Z \mu}+\boldsymbol{\psi}+\boldsymbol{\epsilon},\qquad \text{and} \qquad \boldsymbol{\psi} \sim GP(0,\tau^2\boldsymbol{C}_\phi),\\
 &\boldsymbol{g}(\boldsymbol{X})=\boldsymbol{B\theta}=\sum_{l=1}^p\boldsymbol{B_l\theta_l}= \sum_{l=1}^p\sum_{k=1}^{K_l}B_{kl}\theta_{kl},\\
 \end{aligned}
\end{gather}
where ${B_{kl}}$ is the $k^{th}$ basis functions for a given set of $K_l$ knots for the $l^{th}$ covariate $\boldsymbol{x_l}$, $\theta_{kl}$ is the corresponding basis coefficient, and the spatial random effect $\boldsymbol{\psi}$ follows a Gaussian Process with covariance kernel $\boldsymbol{C}_\phi$. 
\paragraph*{Covariate effects.} 
 We decompose our covariate effects in equation (\ref{equ_pop}) into global effects ($\boldsymbol{g}(\boldsymbol{X})$) and patient-specific intercepts ($\boldsymbol{Z \mu}$). Patient specific intercepts, $\boldsymbol{\mu}$, allow us to account for the variability between patients that can obscure the global impact of $\boldsymbol{X}$ across patients. The associations between the $\boldsymbol{X}$ and $\boldsymbol{y}$ across all patients may be complex and require a flexible solution to estimate. \texttt{MoSAIC} accommodates possible non-linear associations and smooths across patients by specifying spline-based basis functions (\citealt{splineboor,splineDierckx}). Specifically, we use penalized splines as specified by \cite{Eilers}; relevant alternatives include Bayesian penalized splines (\citealt{pspline}) and low-rank predictive processes (\citealt{lowrank}). Thus, we estimate a single set of coefficients $\boldsymbol{\theta}$ across patients and individualize tumor gradients for each patient through $\boldsymbol{B}$. 

We assign a normal prior to the patient intercepts, $\boldsymbol{\mu} \sim N(\boldsymbol{0},\sigma_Z^2\boldsymbol{I}_{I})$, and covariate estimates, $\boldsymbol{\theta_l} \sim N(\boldsymbol{0},\sigma^2_l\boldsymbol{K})$, for all $l = 1,\dots p$. $\boldsymbol{K}$ is a positive-definite matrix, and we set $\boldsymbol{K}=\boldsymbol{I}$ when $\boldsymbol{g}(\boldsymbol{x_l})$ is linear and $\boldsymbol{K}=\boldsymbol{D^TD}$ when $\boldsymbol{g}(\boldsymbol{x_l})$ is non-linear. The penalty matrix $\boldsymbol{D}$ is the second-order difference operator matrix that shrinks the difference between neighboring basis coefficients, leading to smoothed estimation of the covariate effects (\citealt{Eilers}). The hyperparameters $\{\sigma_1^2,\dots,\sigma_p^2\}$ control the amount of shrinkage on $\boldsymbol{\theta}=\{\boldsymbol{\theta_1}^T,\dots,\boldsymbol{\theta_p}^T\}^T$ (towards $0$), further enabling smoothing as well as controlling overfitting. If we standardize our covariates to be on a similar scale, we can apply the same $\sigma_X^2$ for all $p$ covariates. For linear terms, $\sigma^2_l$ is set to a large value (minimal shrinkage); otherwise, we apply an inverse-gamma priors to both variance terms ${\sigma^2_X} \sim IG(a_x,b_x)$ and $ \sigma^2_Z \sim IG(a_z,b_z)$, allowing adaptive shrinkage. 
\paragraph*{Spatial random effects.}
The spatial random effect, $ \boldsymbol{\psi}$, captures the spatial relationship between FOVs within a given tumor tissue, $\mathcal{S}_i$. We expect FOVs in close spatial proximity to be more similar than those that are more distant within a given tumor tissue. Thus, our GP assumption on $ \boldsymbol{\psi}$ yields $\boldsymbol{\psi} \sim N(\boldsymbol{0},\tau^2\boldsymbol{C}_\phi)$, where $\boldsymbol{C}_\phi$ is the kernel matrix. In our RCC setting, $\boldsymbol{C}_\phi$ specifies the correlation between two FOVs as a function of their distance and the spatial decay term, $\phi$. The spatial decay $\phi$ controls the degree of spatial smoothing between FOVs with large $\phi$ values limiting smoothing to FOVs very near one another. Importantly, we only want to smooth among locations \textit{within the same patient} since observations from two different patients do not have a scientifically meaningful spatial relationship. Therefore, we specify the elements of  $\boldsymbol{C}_\phi$ via a squared exponential kernel such that there is independence across patients:
\begin{align*}
  {C}_\phi(s_{i},s'_{i'})&=exp\{-\phi \lVert s_{i}-s'_{i'}\rVert^2\}\mathbb{I}(i=i').
\end{align*} 
The parameter $\tau^2$ captures the spatial variance magnitude. We borrow information across patients to estimate common $\tau^2$ and $\phi$ since we expect there to be a similar spatial structure across patients. We posit a non-informative inverse-gamma prior on $\tau^2$, such that $\tau^2 \sim IG(a_s,b_s)$. We fix the spatial decay hyperparameter, $\phi$, from a series of values through a test-training procedure (details in Supplement S1).
\paragraph*{Measurement error.}
 The measurement error term $\boldsymbol{\epsilon}$ captures any remaining variability in the outcome not explained by the covariate gradient, spatial relationship between observations, or patient differences. We specify $\boldsymbol{\epsilon} \sim N(\boldsymbol{0},\sigma_y^2\boldsymbol{I}_{N})$ and $\sigma^2_y \sim IG(a_y,b_y)$.
\subsection{Posterior estimation and inference}
\paragraph*{Posterior sampling.} Since we posit Gaussian priors to $\boldsymbol{\theta}, \boldsymbol{\psi},\boldsymbol{\mu} \text{ and } \boldsymbol{\epsilon}$, we can sample from the marginal model, 
\begin{gather}\label{marginalequ}
 \begin{aligned}
 \boldsymbol{y} &\sim N(\boldsymbol{0}, \boldsymbol{\Sigma_y}); \quad \boldsymbol{\Sigma_y} =\sigma_X^2\sum_{l=1}^p\boldsymbol{B_l}\boldsymbol{K}\boldsymbol{B_l^T}+{\sigma^2_Z}\boldsymbol{Z}\boldsymbol{Z}^T+\tau^2\boldsymbol{C}_\phi\boldsymbol{}+\sigma^2_y\boldsymbol{I_N},
 \end{aligned}
\end{gather} 
with $4$ free parameters: $ \sigma^2_Z,{\sigma^2_X},\tau^2, \sigma^2_y$. Marginalized parameters ($\boldsymbol{\theta},\boldsymbol{\mu},\boldsymbol{\psi}$)  are recovered from the joint posterior. This formulation improves convergence and mitigates identifiability concerns between $\boldsymbol{\mu}$ and $\boldsymbol{\psi}$. 
To fit \texttt{MoSAIC}, we implement a robust adaptive Metropolis algorithm as specified by \cite{robamcmc} to sample the variance terms,\{$ \sigma^2_Z,{\sigma^2_X},\tau^2, \sigma^2_y$\}, from equation \ref{marginalequ} (details in Supplement S1). By implementing an adaptive algorithm that updates the proposal distribution to shrink or expand towards the desired acceptance probability, \texttt{MoSAIC} explores the posterior more efficiently. We calculate the Deviance Information Criteria (DIC; \citealt{DICog}) and scaled Watanabe–Akaike Information Criterion (WAIC; \citealt{watanabe2010asymptotic}), which use log predictive density penalized by the number of effective parameters to derive Bayesian measures of model fit (\citealt{Gelman2014_DIC}).
\paragraph*{Joint credible bands and global significance.}
 In addition to estimating the global covariate effects $\boldsymbol{g}(\boldsymbol{X})$, we are interested in quantifying the amount of uncertainty around those estimates and testing for significance. While point-wise credible bands suffice for linear coefficients, there is an inherent multiple testing issue when inferring smoothed covariate functions since we are generating estimates over a continuum of values $\boldsymbol{X}$ (\citealt{RuppertWandCarroll2009}). To address this, we use joint credible bands similar to those proposed by \cite{jointbands}. Briefly, we estimate $100 (1-\alpha)^{th}$ joint-credible bands such that the probability of $\boldsymbol{g}(\boldsymbol{X})$ being in that interval is greater than or equal to $1-\alpha$ for \textit{all values} of $\boldsymbol{X}$, thus naturally allowing for multiple testing correction. This strategy also enables assessment of a global Bayesian p-value to test the null hypothesis that the relationship between our tumor gradient and the outcome is significantly different from $0$ across the entire tumor gradient (further details in Supplement Section S1).
 
\paragraph*{Variance decomposition.}
The construction of the \texttt{MoSAIC} model can be used to delineate the influence of each source of heterogeneity on the outcome. Given the marginal model from equation \ref{marginalequ}, we can write the joint covariance of our outcomes ($\boldsymbol{\hat{\Sigma}}_Y$) as the summation of four matrices:
$\boldsymbol{\hat{\Sigma}}_Y=\boldsymbol{\hat{\Sigma}}_X+\boldsymbol{\hat{\Sigma}}_Z+\boldsymbol{\hat{\Sigma}}_{\psi}+\boldsymbol{\hat{\Sigma}}_{\epsilon}$.
Thus, the variability in our outcome explained by the tumor gradient ($\boldsymbol{\hat{\Sigma}}_X$), patient differences ($\boldsymbol{\hat{\Sigma}}_Z$), the spatial relationship between FOVs ($\boldsymbol{\hat{\Sigma}}_{\psi}$), and the measurement error ($\boldsymbol{\hat{\Sigma}}_{\epsilon}$). Analogous to $R^2$ in ANOVA and regression (\citealt{ANOVA}; \citealt{GLMM_R2}), we can find the variance contribution attributable to each source of heterogeneity using the percentage of variance explained (PVE) as: 
\begin{align*}
    PVE[\boldsymbol{\Sigma}_{(\cdot)}] &= \frac{trace[\boldsymbol{\Sigma}_{(\cdot)}]}{trace[\boldsymbol{\Sigma}_Y]}\times100.
\end{align*}
We estimate the traces for $\boldsymbol{{\Sigma}}_{X},\boldsymbol{{\Sigma}}_Z,\text{ and }\boldsymbol{{\Sigma}}_{\psi}$ from the empirical variance of the MCMC samples of $ \boldsymbol{B\hat{\theta}}, \boldsymbol{Z \hat{\mu}}, \text{ and }\boldsymbol{\hat{\psi}}$, respectively. $\boldsymbol{{\Sigma}}_{\epsilon}$ is estimated by $\bar{\sigma}^2_y\boldsymbol{I_N}$, where $\bar{\sigma}^2_y$ is the posterior mean of the measurement error. The PVE allows us to discern relative contribution of the tumor gradient in explaining changes in $\boldsymbol{y}$ compared to the between-patient heterogeneity and the spatial relationships between the FOVs.
\section{Simulations}\label{sec-sim}
\begin{figure}[ht!] \centering
 \includegraphics[width=1\linewidth]{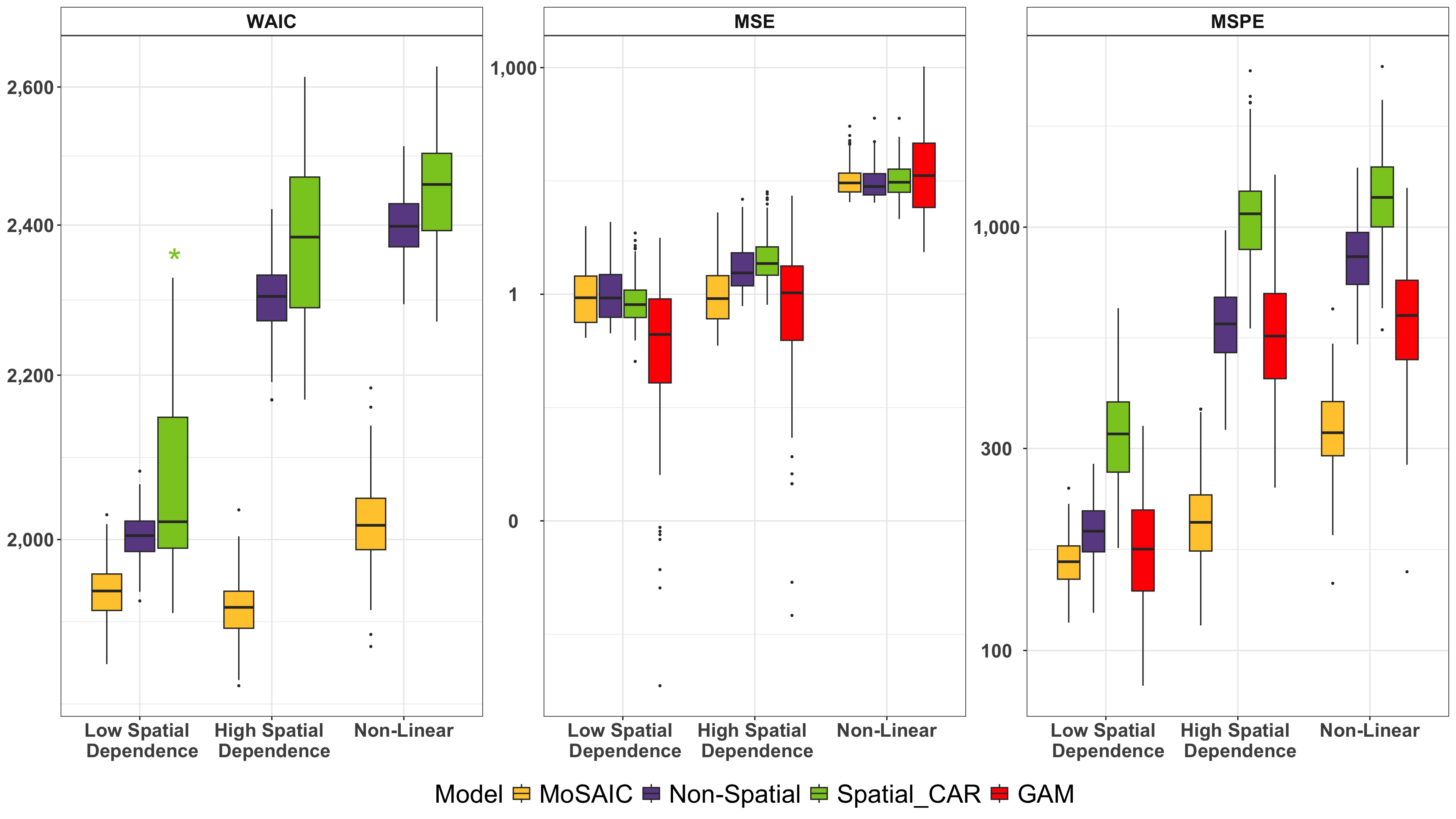}
 \caption{\textbf{Simulation performance of \texttt{MoSAIC}}. Each box plot is colored to indicate which of the four models it represents. The left panel plots the WAIC values for Bayesian models to assess model fit. The middle panel plots the MSE of $g(\boldsymbol{X})$ to assess estimation accuracy. The right panel plots the MSPE to assess models predictive accuracy. The y axis on the log base 10 scale. Three WAIC values for the Spatial\_CAR model in Scenario 1 were less than 1,000 and removed for clarity as indicated by (\textcolor{green}{$\star$}).}
 \label{fig:sim}
\end{figure}

To demonstrate the effectiveness of \texttt{MoSAIC}, we conduct replicated simulation studies via ablation experiments to evaluate performance in terms of estimation accuracy, model fit, and prediction. We begin by outlining the HMI data generation process, followed by descriptions of the competing methods and comparative metrics. We assess the performance of \texttt{MoSAIC} under three scenarios: low spatial dependence 
(Scenario 1), high spatial dependence (Scenario 2) and non-linear tumor gradient effects (Scenario 3). All results are aggregated over 200 independent replicates
\paragraph*{Simulation design.} 
 Across all scenarios, we generate $200$ data sets with $300$ FOVs distributed unequally among $20$ patients such that $\boldsymbol{y}=\boldsymbol{\mu}+g(\boldsymbol{X})+\boldsymbol{\psi}+\boldsymbol{\epsilon}$. We set $\sigma^2_y=50$ and sample our patient-specific intercepts such that $\boldsymbol{\mu} \sim N(\boldsymbol{50},2\sigma^2_y\boldsymbol{I}_{20})$. In Scenarios 1 and 2, we assume a linear covariate effect ($g (\boldsymbol{X})=\boldsymbol{X}\theta_1$) while in Scenario 3, we specify a non-linear effect, $g (\boldsymbol{X})=\rm{arctan} (\boldsymbol{X})\theta_1$. In the low spatial dependence setting (Scenario $1$), the measurement error $\sigma^2_y$ and the spatial variance $\tau^2$ are equal, so that between-patient differences account for more of the variability in the outcomes than the spatial structure. In Scenario $2$, $\tau^2$ is ten times larger than $\sigma^2_y$, so now spatial dependence dominates outcome variability relative to between-patient differences. We also decrease the spatial decay parameter to reflect a weaker and stronger spatial coherence across FOVs in scenarios $1$ and $2$, respectively. Scenario $3$ specifies a spatial dependence strength intermediate between those of Scenarios $1$ and $2$. 
We hypothesize that \texttt{MoSAIC}'s performance relative to currently applied models will improve as the spatial dependency between FOVs increases and in the presence of non-linear covariate effects. 
\paragraph*{Comparative metrics and methods.} We assess estimation accuracy via mean squared error (MSE) of the covariate curves $g (\cdot)$ and model fit for through rescaled WAIC (\citealt{Gelman2014_DIC}). To evaluate predictive accuracy, $30$ FOVs ($10$\%) are withheld as a test dataset in each simulation, enabling calculation of the mean squared predictive error (MSPE) and the $95$\% predictive coverage of both $\boldsymbol{y}$ (details in Supplement Section S2). We compare these metrics across four models: \texttt{MoSAIC}, a non spatial version of \texttt{MoSAIC} without a spatial random effect, $\boldsymbol{\psi}$, in equation \ref{equ_pop} (Non-Spatial), a generalized additive model with random effects for patients (GAM; \citealt{gam}), and a spatial regression fit with a conditional autoregressive (CAR) spatial random effect (Spatial\_CAR). We implement the Spatial\_CAR model specified by \cite{Leroux1999} through the \texttt{CARBayes} package by \cite{duncancar}. We use R package \texttt{gam} to implement the GAM model (\citealt{gam_r}). We can recover samples of $\boldsymbol{\mu, \theta,\psi}$ by sampling from their posteriors to be used later for prediction (details in Supplement Section S2). We specify that $\boldsymbol{\psi}$ has a mean of $0$ within a tumor tissue, $\frac{1}{n_i}\sum_{j=1}^{n_i}\psi(s_{j(i)})=0 \quad \text{for all } i$, to improve identifiability between $\boldsymbol{\psi}$ and $\boldsymbol{\mu}$. For the non-linear simulation scenario, we use the R package \texttt{mgcv} (\citealt{rbasesfunction}) to form the splines for \texttt{MoSAIC} and Non-Spatial models. We use the \texttt{bs} function by \cite{BSfunction} for the Spatial\_CAR model and smoothing splined with 5 degrees of freedom for the GAM. We use the \texttt{ramcmc} package to update the proposal covariance (\citealt{ramcmc}). We run $60,000$ MCMC iterations with an adaptive phase of $30,000$ iterations and total burn-in of $45,000$.
\paragraph*{Results.}
 Figure \ref{fig:sim} summarizes the performance of each model across all three scenarios, where each box plot displays the distribution of each metric over the $200$ simulated datasets, colored by model. Across all three simulation scenarios, \texttt{MoSAIC} achieves lower (better) WAIC, lower MSPE, and closer-to-nominal $95\%$ predictive coverage than all competing models, with comparable or lower MSE. The left-most panel shows the distribution of WAIC values, with \texttt{MoSAIC} consistently achieving lower values, indicating a better model fit to the data. 
The center panel of Figure \ref{fig:sim} shows how the MSE values of $g(\boldsymbol{X})$ across all 200 simulations, with lower values indicating improved estimation accuracy. The GAM model achieves the lowest MSE in the low spatial dependence setting. However, its performance declines as the spatial dependency and the covariate curve complexity increase. In Scenarios 1 and 2, \texttt{MoSAIC} has lower MSE than the Non-Spatial and Spatial\_CAR models. In the non-linear setting, \texttt{MoSAIC} produces slightly higher MSE than the Non-Spatial model (15\% higher), though the fitted curves are often visually comparable (Supplement Figure S1). 
The small losses in efficiency of \texttt{MoSAIC} are countered by the considerable gains in predictive accuracy, as shown by the distribution of MSPE in the right most box-plot. 
 For example, \texttt{MoSAIC} has an MSPE that is $15$\%, $64$\%, and $60$\% lower than the Non-Spatial model in Scenarios 1, 2, and 3, respectively. 
 Moreover, the average predictive coverage for \texttt{MoSAIC} is about $95$\% in the linear scenarios and increases from $95.3$\% for the point-wise credible bands to $99$\% for the joint credible bands in the non-linear scenario. Additional simulation details can be found in Supplement Table S1.

In summary, \texttt{MoSAIC} demonstrates consistently strong and often superior performance across varying degrees of spatial dependence and covariate function complexity relative to currently available methods applied to HMI data. The substantial gains in model fit and predictive accuracy as spatial dependence increases confirm the importance of jointly modeling patient-level and spatial random effects in this data context, and support \texttt{MoSAIC} as a methodologically principled choice for the analysis of multi-resolution spatial data.

\section{Spatial Tumor Gradients in Renal Cell Carcinoma}\label{sec-rd}
\begin{figure}[ht!] \centering
 \centering
 \includegraphics[width=1\linewidth]{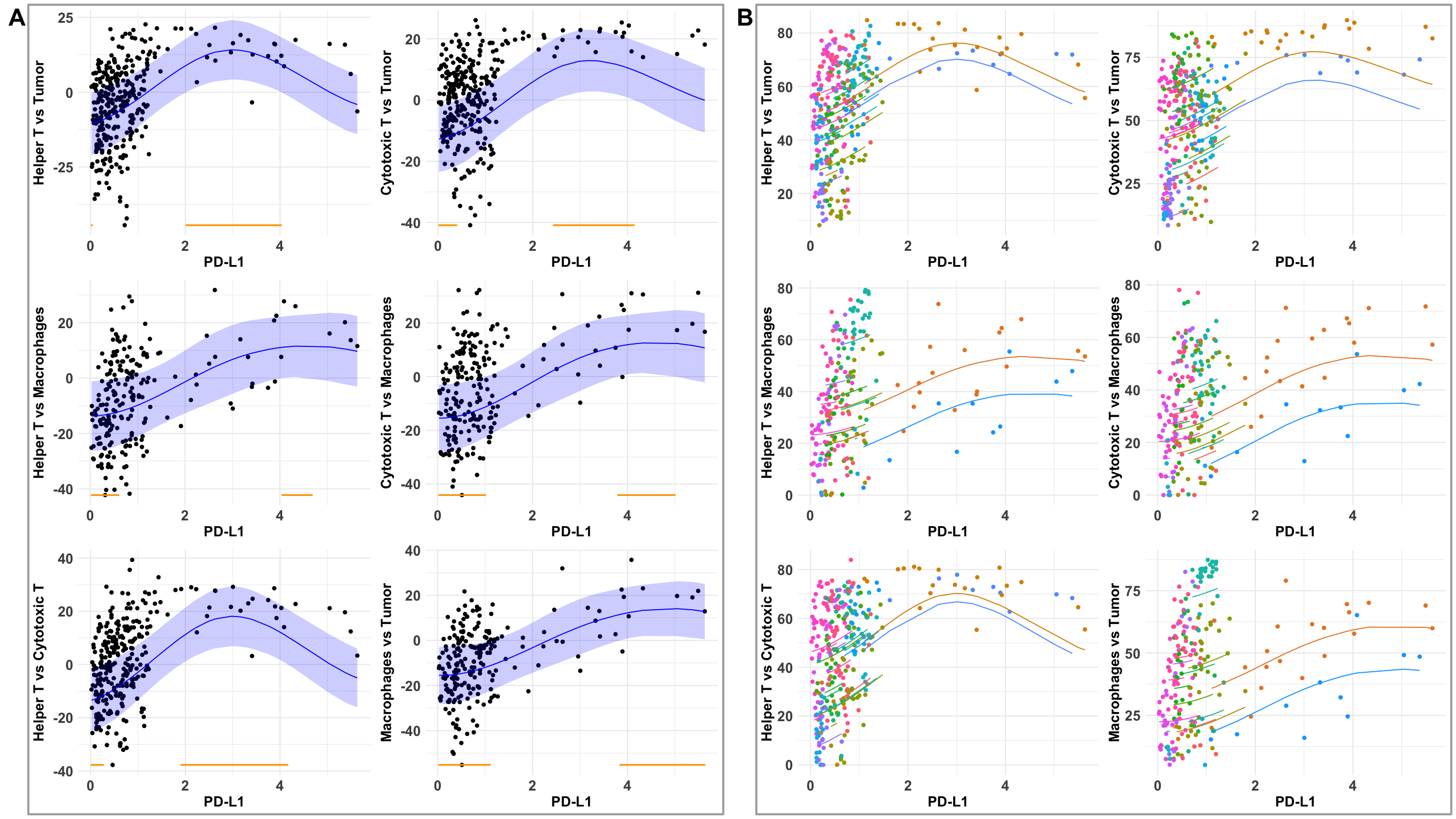}
 \caption{ \textbf{A. Global curve for the associations between PD-L1 and pairwise cellular colocalizations with joint credible bands.} The y-axis is the cell colocalization centered at the patient intercept. Orange lines indicate locations where our curve significantly deviates from $0$ and blue shaded ribbons display the joint credible bands. \textbf{B. Patient-specific curves for the association between PD-L1 and pairwise cellular colocalizations}. Each color and curve represents a unique patient ID. }
 \label{fig:PDL1}
\end{figure}

Our motivating RCC HMI dataset arises from a study of $N=354$ FOVs across $I=21$ patient tumor tissues with both clear cell RCC and sRCC features following established imaging protocols (\citealt{mIFmethod}\anonflag{; \citealt{MayPrePrint}}). Specifically, Vectra® Polaris™ Work Station and inForm® Cell Analysis™ software (Akoya Biosciences) were used to obtain images and quantify marker expressions, respectively. The resulting quantification of $3$ phenotypic markers (CD4, CD8, and CD163) defines $4$ cell types: helper T, cytotoxic T, macrophages, and Tumor (see Supplemental Table S2). On average, tumor cells are the most abundant, comprising over $90$\% of the total cells within an FOV, while macrophages are the rarest occurring cell at $0.1$\% of total cells within an FOV. We quantify cellular colocalization using a distance-based metric DIMPLE by \cite{MASOTTI2023100879}, $\boldsymbol{y}_{\text{DIMPLE}}$, rescaled such that $0$ indicates complete separation (no colocalization) and $100$ indicates perfect spatial overlap (high colocalization) in each FOV (see Supplement Section S3). Figure \ref{histogram} shows the distribution of $\boldsymbol{y}_{\text{DIMPLE}}$ for each pair of cells, demonstrating the variability in cell colocalizations across both FOVs and cell pairs. 
\begin{figure}[ht!] 
    \centering
    \includegraphics[width=1\linewidth]{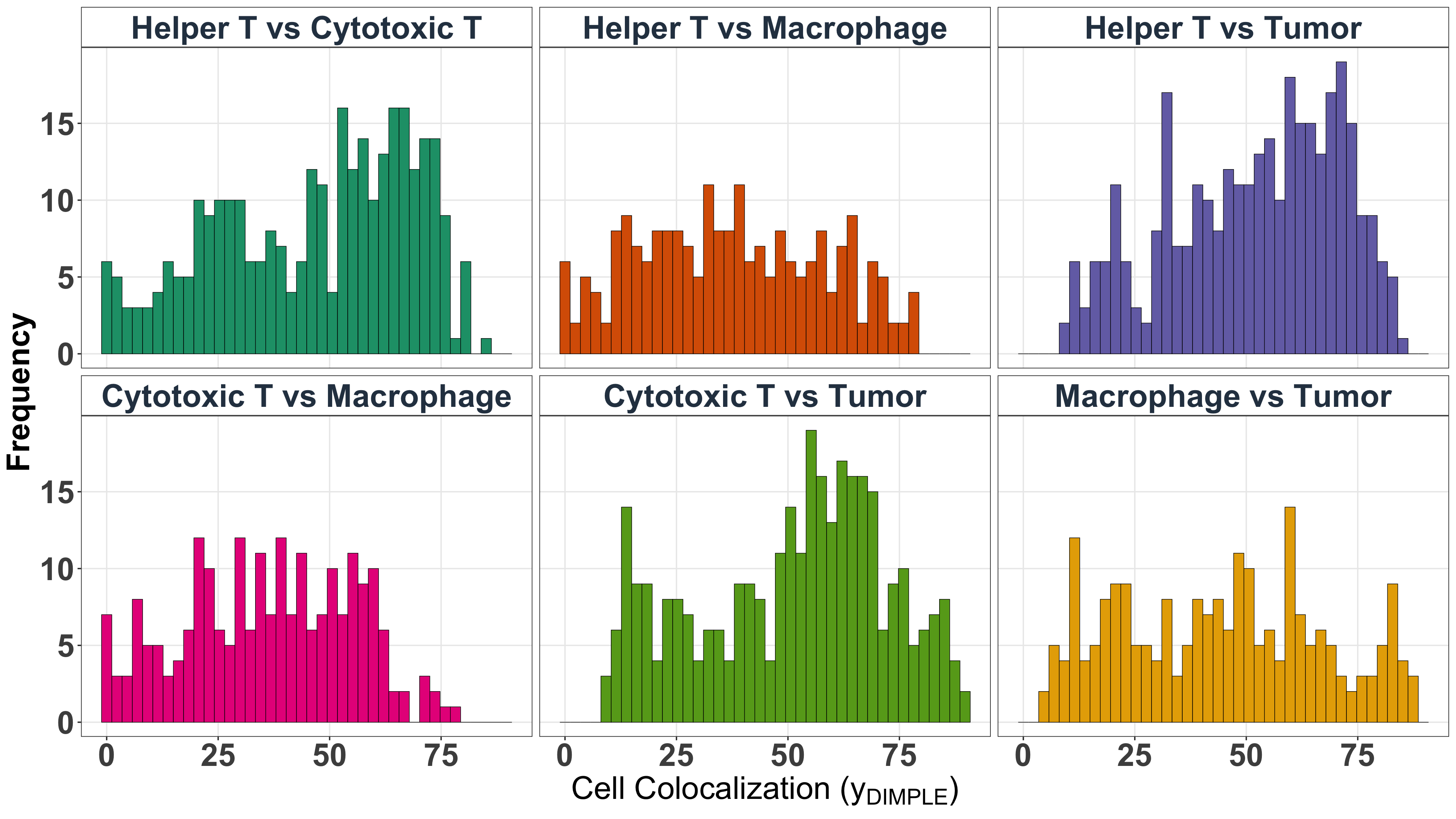}
    \caption{\textbf{ Distribution of cell colocalization within FOVs across pairwise comparisons.} Each panel shows the distribution of cell colocalization scores across FOVs for each of the six pair of cells. The x-axis is the adjusted DIMPLE distance colocalization score and the y-axis is the number of FOVs with that colocalization score across all $21$ patients.}
    \label{histogram}
\end{figure}
Two functional biomarker expressions define the tumor gradient at the FOV level: Programmed Death Ligand 1 (PD-L1) and N-cadherin (NCAD). PD-L1, which is a common target for cancer immunotherapies, has increased expression in sarcomatoid dedifferentiated tumors (\citealt{PDL1}) and NCAD is typically expressed at higher levels in more mesenchymal-like cells (\citealt{NCADdef}). Thus, higher NCAD and PD-L1 expression indicates more advanced progression along the EMT gradient toward sRCC. Since our outcomes are measured at the FOV level, we summarize cellular NCAD and PD-L1 expression across FOVs (details in Supplement Section S3). Using \texttt{MoSAIC}, we investigate (a) the global, across patients relationships between cell colocalizations and the tumor gradients defined by PD-L1 and NCAD, and (b) how the spatial gradients between FOVs differ among the six cell-type pairs.\\
We implement $6$ independent models, one for each pairwise cell colocalization, as follows: 
\begin{gather}\label{equ:real}
 \begin{aligned}
 \boldsymbol{y}_{\text{DIMPLE}}= & \boldsymbol{g(\{\textbf{PD-L1}, \textbf{NCAD}\})}+\boldsymbol{Z \mu}+\boldsymbol{\psi}+\boldsymbol{\epsilon}.\\
 \end{aligned}
\end{gather}
We use penalized splines with $5$ knots and a common shrinkage ($\sigma^2_X$) to capture the tumor gradient effect of PD-L1 and NCAD on the colocalization between a pair of cells. Each model includes FOVs with at least $1$ of each cell type, with all models having a sample size between $215$ and $325$. We run \texttt{MoSAIC} for $60,000$ MCMC iterations with a burn-in of $45,000$, yielding $15,000$ posterior samples in $4.4$ to $9.7$ minutes on a laptop with an M1 chip and 16GB of memory. We determine model convergence through trace plots of MCMC samples and by using Geweke diagnostic scores (\citealt{Geweke.diag}; see Supplement Figure S2 and Table S3). Our results are summarized as follows: global and patient specific PD-L1 and NCAD effects, the percentage of variance explained by each component in Equation \ref{equ:real}, and the variability in spatial gradients across models.
 \paragraph*{PD-L1.} The global associations between PD-L1 expression and the six pairwise cell colocalizations are shown in Figure \ref{fig:PDL1}A. Our results show two unique patterns of associations. Most associations display an increasing pattern in colocalizations with PD-L1 expression initially, suggesting a potential increase in immune activity. As indicated by the orange lines on the x-axis, macrophages continue to co-locate significantly closer to all other cells as PD-L1 increases (middle row and bottom-right). The other three curves eventually return to $0$ (top row and bottom-left). PD-L1's increasing relationship with macrophages and tumor cell colocalization (bottom-right) and the non-monotonic relationship with cytotoxic T cells and tumor cell colocalization (top left) may explain why sRCC has a worse prognosis and increased susceptible to immunotherapy overall. Increased PD-L1 expression is associated with increased tumor-associated macrophages, T cell inhibition, and worse prognosis (\citealt{TAM_ref}). Thus, therapies that inhibit PD-L1 may improve T cell tumor infiltration and address this attenuated immune response to the tumor. Figure \ref{fig:PDL1}B shows patient specific curves, which highlights the noticeable between patient variability. Much of the limited data at higher PD-L1 expressions belonging to only two patients. 
 
\begin{figure}[ht!] 
 \centering
 \includegraphics[width=1\linewidth]{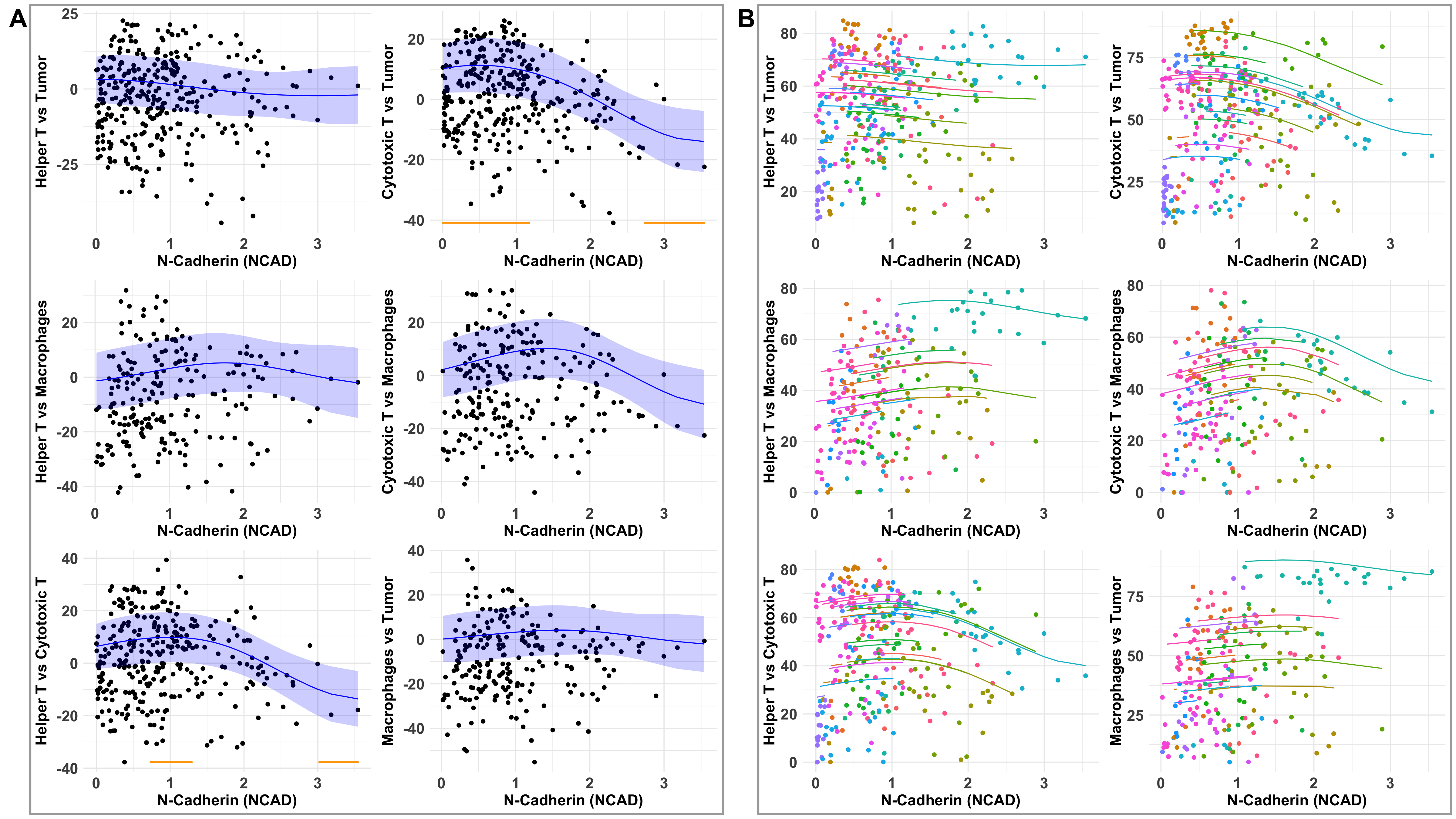}
 \caption{\textbf{A. Global curve for the associations between NCAD and pairwise cellular colocalizations}. The y-axis is the cell colocalization centered at the patient intercept. Orange lines indicate locations where our curve significantly deviates from 0 and blue shaded ribbons display the joint credible bands. \textbf{B. Patient-specific curves for the association between NCAD and pairwise cellular colocalizations.} Each color and curve represents a unique patient ID.}
 \label{fig:NCAD}
\end{figure}

\paragraph*{NCAD.} Figure \ref{fig:NCAD}A shows the analogous global tumor gradient associations between NCAD expression and the six pairwise cell colocalizations. We see significant associations between cytotoxic and Tumor (top-right) and helper T vs cytotoxic T (bottom-left) colocalizations with NCAD (marked in orange on the x-axis). In both instances, lower values of NCAD indicate higher colocalization with a decreasing pattern as the expression of NCAD increases. This finding of increased cellular engagement between cytotoxic T cells and tumor cells at the lower end of the EMT gradient is consistent with previous findings (\citealt{MayPrePrint}) and suggests poorer immune targeting of tumor cells in sRCC tissue. Since cytotoxic T cells are primarily responsible for tumor cell destruction, less colocalization between cytotoxic T and tumor cells may suggest lower infiltration in higher EMT gradients (more mesenchymal regions). Moreover, \cite{MayPrePrint} found that visually, cytotoxic T cells tend to cluster near the boundary of sRCC, which may explain the change from initially higher to lower colocalization between these cells as NCAD increases. Figure \ref{fig:NCAD}B shows the corresponding patient-specific curves and their deviations from global effects. Analogous to PD-L1, there is considerable inter-patient heterogeneity, primarily stemming from differing numbers of FOVs among the patients.

\begin{figure}[ht!]
    \centering  \includegraphics[width=1\linewidth]{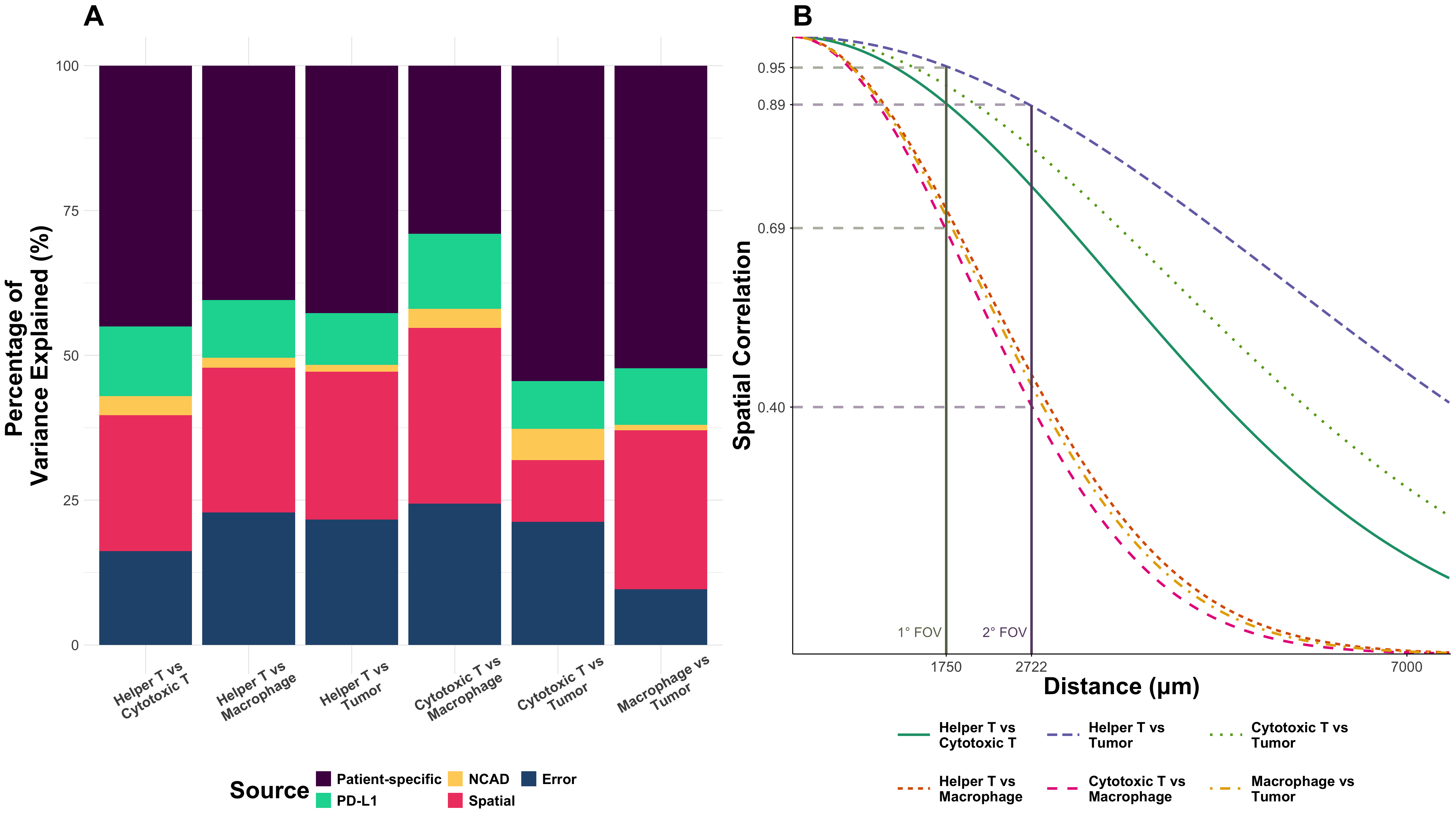}
    \caption{\textbf{A. The percentage of variance explained by each component of the \texttt{MoSAIC} regression model.} Each bar is a pairwise cell colocalization model. Each color represents the relative contribution of each a source of variability to the total variance. Top: patient-specific (purple)- PD-L1 (green)-NCAD(yellow)-spatial(red)-measurement error (blue):Bottom. 
    \textbf{B. Spatial Gradients across the six pairwise cell-colocalization models.} Curves illustrate the spatial correlation between FOVs at a given distance. Vertical lines indicate the average distance between first- and second-order neighbors.}
    \label{fig:variance}
\end{figure}

\paragraph*{Percentage of Variance Explained and Spatial Gradients.}
Figure \ref{fig:variance}A shows the PVE across the 6 models from different sources, namely patient-specific, global covariate effects (PD-L1 and NCAD), spatial random effects, and measurement error. Overall, our MoSAIC model is able to explain at least $75$\% and up to $90$\% of the total variability in our outcomes with the rest attributed to measurement error. Congruent to Figures \ref{fig:PDL1} and \ref{fig:NCAD}, the majority of the variance in our outcomes is captured by the patient-specific effects, accounting for 29\% to 54\%, and PD-L1 accounts for more of the variance as compared to NCAD across all models (8\%-13\% vs <1\%-5\%). Finally, the percentage of spatial variability between FOVs varies across models, ranging from 11\% to 30\%, indicating the existence of non-negligible spatial variation accounted for by \texttt{MoSAIC}. 

To further explore the spatial gradients, Figure \ref{fig:variance}B summarizes the spatial correlations across each of the six models. The curves show the decrease in spatial correlation of the cell colocalization measures between two FOVs as a function of their distance. Vertical lines indicate the average distance between first and second order FOV neighbors across the $21$ tumor tissues, with high to moderate spatial gradients. The helper T-tumor cell colocalizations had the most coordinated spatial gradient, with the correlation between FOVs decreasing slightly from $0.95$ to $0.89$ between first- and second-order FOV neighbors. The spatial gradient is less smooth for cytotoxic T-macrophage colocalization, with the correlation dropping from $0.69$ to $0.40$ between first- and second-order neighbors. 

We further ran a goodness of fit test comparing the WAIC and DIC of \texttt{MoSAIC} to the non-spatial alternative presented in Supplement Table S4. \texttt{MoSAIC} fits the data better in each of the 6 models, further supporting the inclusion of spatial random effects between FOVs in our analysis. 

\section{Discussion} \label{sec-conc}
We propose a novel hierarchical Bayesian spatial regression model, \texttt{MoSAIC}, for multi-resolution spatial imaging data, which complements current scientific understanding of spatial heterogeneity of TME in cancer. By explicitly delineating sources of variability and borrowing information across replicated spatial surfaces, \texttt{MoSAIC} reveals global biological trends, estimates tumor gradients, and characterizes their associations with the TME. Simulation studies demonstrate that \texttt{MoSAIC} achieves superior prediction accuracy and model fit relative to existing spatial and non-spatial alternatives, with performance gains most pronounced in settings with strong spatial dependence and complex tumor gradients.

Motivated by and applied to an sRCC HMI dataset, \texttt{MoSAIC} identifies that increased PD-L1 expression is associated with greater macrophage and tumor cell colocalization, consistent with the established literature linking elevated PD-L1 to an increase in tumor-associated macrophages. The initially increasing colocalization of T cells with tumor cells as PD-L1 expression rises may further indicate partial immune infiltration of sRCC that is functionally suppressed until PD-L1 is inhibited, offering a mechanistic basis for the heightened immunotherapy susceptibility observed in sRCC. Additionally, cytotoxic T cell–tumor cell colocalization decreases with increasing NCAD expression, suggesting lower immune targeting in more mesenchymal regions. This change in cell colocalization could explain the poorer prognosis of sRCC, suggesting a potentially reduced capacity of the immune system to target tumor cells in sRCC regions. Our model captures at least 75\% of total variance in pairwise cellular colocalizations models, demonstrating strong explanatory power across multiple resolutions of heterogeneity. Generally, patient specific effects account for the largest PVE, though spatial relationships between FOVs accounting for $10\%-30\%$. This, in addition to the substantial improvement in fit over non-spatial models, underscores the importance of explicitly modeling spatial structure in HMI data and affirms \texttt{MoSAIC}'s profitability in such settings. Beyond its inferential capabilities, \texttt{MoSAIC} is computationally efficient, with each model fit within minutes on a standard laptop (see Supplementary Table S1). In summary, \texttt{MoSAIC} offers a unified investigation of the TME across patients, incorporating several layers of heterogeneity originating from spatial dependence and between- and within-patient variability. 

 \texttt{MoSAIC} is broadly applicable to hierarchical or multi-resolution spatial imaging data for assessing tumor gradients across patient cohorts, and several natural extensions would further enhance its scope. Incorporating patient-level covariates, such as treatment response or cancer stage, would attribute additional variance to clinically relevant sources and incorporate yet another resolution. Currently, we model all pairwise cellular colocalizations independently, which allows us to investigate colocalizations with very different spatial gradients as shown in Figure \ref{fig:variance}B. Joint modeling of multiple pairwise cellular colocalizations would enable exploration of multi-cellular interactions. For example, one could characterize the interplay among T cells, macrophages, and tumor cells along the PD-L1 gradient. Finally, \texttt{MoSAIC} can be extended to model higher-resolution cellular level data; however, doing so requires nontrivial methodological development to accommodate the substantial increase in dimensionality, from hundreds of FOVs to millions of cells. Developing scalable methods for this setting is an important direction for future research.
 
As cancer medicine advances toward increasingly personalized approaches, there is a growing need for rigorous statistical frameworks capable of integrating the multi-scale, spatially structured data generated by modern imaging technologies. Multiplex imaging and related platforms provide an exceptionally rich source of biological information, but realizing their full scientific potential demands models that are both methodologically principled and scientifically motivated. \texttt{MoSAIC} addresses this need directly, filling a critical methodological gap by enabling quantitative, multi-resolution investigation of the spatial processes governing tumor biology. Its capacity to identify potential biomarker targets within the tumor gradient positions \texttt{MoSAIC} as a valuable tool for advancing immunotherapy development and, more broadly, for deepening our quantitative understanding of the microenvironmental drivers of patient outcomes. 
\section{Acknowledgments}\label{disclosure-statement}
\anonflag{
JA's work was supported by the Rogel Cancer Center Graduate Student Scholarships of the University of Michigan Rogel Cancer Center and was partially supported by National Institutes of Health grant CA $83654$. VB’s work was supported by the National Institutes of Health grants R01CA244845-01A1 and P30 CA46592 and funds from the University of Michigan Rogel Cancer Center and School of Public Health
}
GPT 5.2 Generative AI model (\citealt{GPT_52}) was used in accordance with journal guidelines to refine the grammar/tone of the paper, format BibTeX citations, and refine the elements of the latex displays (algorithms and tables). Claude Sonnet 4.6 (\citealt{Claude}) and GPT 5.2 were used to annotate and streamline R scripts for public access on Github\anonflag{ repository \href{https://github.com/jcaldous/MoSAIC}{jcaldous/MoSAIC}}. All outputs were reviewed and revised to display accurate information. \anonflag{These tools were accessed through the University of Michigan GenAI Services (\citealt{umich_genai_2025}) and Visual Studio Codes' Github co-pilot extension (\citealt{github_copilot_2025}).}

\anonflag{
\section{Data Availability Statement}\label{data-availability-statement}
Data will be available with the publication of \cite{MayPrePrint}, whose preprint can be found on bioRxiv. Multiplex immunofluorescent data will be shared by the lead contact, Dr. Evan Keller, upon request at etkeller@umich.edu.
}
\newpage 
 \bibliography{bibliography.bib}

\newpage

\phantomsection\label{supplementary-material}
\bigskip

\begin{center}

{\large\bf Supplementary Materials for Multi-Resolution Spatial Regression Analysis of Cellular Colocalizations in Cancer Imaging }

\end{center}
\makeatletter
\setcounter{figure}{0}
\setcounter{table}{0}
\setcounter{equation}{0}
\renewcommand \thesection{S\@arabic\c@section}
\renewcommand \thetable{S\@arabic\c@table}
\renewcommand \thefigure{S\@arabic\c@figure}
\renewcommand \thealgorithm{S\@arabic\c@algorithm}
\renewcommand \theequation{S\@arabic\c@equation}
\renewcommand{\labelenumi}{S\arabic{enumi}.}
\renewcommand{\labelenumii}{S\arabic{enumi}.\arabic{enumii}.}
\makeatother
\bigskip
\textbf{Table of Contents.}\label{supp:toc}

\begin{enumerate}
  \item \textbf{Methods}
    \begin{enumerate}
      \item Estimating spatial decay $\phi$
      \item Robust adaptive Metropolis-Hastings algorithm
      \item Joint credible bands and global significance
    \end{enumerate}

  \item \textbf{Simulation}
    \begin{enumerate}
      \item Calculation of Predictive Coverage and Mean Squared Predictive Error
      \item Fitted covariate curves in the non-linear simulations
      \item Recovering $\boldsymbol{\beta}$
      \item Summary Table for Simulation Results
    \end{enumerate}

  \item \textbf{Sarcomatoid Renal Cell Carcinoma Analysis}
    \begin{enumerate}
      \item Tumor gradient and Immune Cell Definitions for Real Data Analysis
      \item Quantifying cell clustering within an FOV
      \item \texttt{MoSAIC} MCMC Convergence and Model Diagnostics
    \end{enumerate}
\end{enumerate}
\newpage
\suppsection{Methods}\label{supp:meth}
\subsuppsection{Estimating spatial decay $\phi$}\label{supp.phi.selection}
\FloatBarrier
Before fitting \texttt{MoSAIC}, we estimate and fix the spatial decay hyperparameter $\phi$. We implement a test and training procedure to select the best $\phi$ from a range of values $\boldsymbol{\phi}_{range}$ as shown in Algorithm \ref{algo_phi}. For each potential value $\tilde{\phi}$, we fit an abbreviated model \ref{equ_pop} with only the spatial random effects and measurement error to  the residuals from the linear regression model, $\boldsymbol{y}^{resid}$. Then, we use the posterior predicted mean for the withheld test data, $\boldsymbol{y}^{resid}_{\text{test}}$, to calculate the root mean squared error. We fix the spatial decay at the value of $\tilde{\phi}$ with the lowest root mean squared error. The test-training function \texttt{phi\_tt\_parallel.R} includes an option to select the spatial decay using a log predictive score, which includes uncertainty calibration in the selection of $\phi$, though we found that both criteria lead to similar selections\anonflag{ (code at \href{https://github.com/jcaldous/MoSAIC}{jcaldous/MoSAIC})}. For our real data analysis we considered spatial decay values from $0$ to $15$ in increments of $0.5$.

\subsuppsection{Robust adaptive Metropolis-Hastings algorithm}\label{supp.MCMC}
To improve convergence, we implement the robust adaptive Metropolis-Hastings procedure specified by \cite{robamcmc}. Let $\boldsymbol{\gamma}$ be a vector of sampled parameters, and $\boldsymbol{S}$ is a lower triangular proposal variance update matrix. Given a target acceptance of 23.5\%, each MCMC sample is derived following Algorithm \ref{algo_mcmc}. This update was implemented through the \texttt{ramcmc} function by \cite{ramcmc} in R.
\subsuppsection{Joint credible bands and global significance}\label{supp.joint}
 For $M$ MCMC samples, let $\boldsymbol{g}^{ (m)}(\boldsymbol{X})$ be a posterior sample from our estimated global curve for a given biomarker expression $\boldsymbol{X}$ within the domain of expression across all patients $\mathcal{X}$. The $100 (1-\alpha)^{th}$ joint-credible bands expand the equivalent point-wise credible bands such that the probability of $\boldsymbol{g}(\boldsymbol{X})$ being in that interval is greater than or equal to $1-\alpha$ for \textit{all values} of $\boldsymbol{X} \in \mathcal{X}$. There are many intervals that satisfy this condition. We construct a $100 (1-\alpha)^{th}$ joint credible band such that, for all values of $\boldsymbol{X} \in \mathcal{X}$ 
\begin{gather}\label{equ_joint}
\begin{aligned}
 I_{\alpha} (s)&= \bar{\boldsymbol{g}}(\boldsymbol{X}) \pm q_{ (1-\alpha)}\hat{Std}\{\bar{\boldsymbol{g}}(\boldsymbol{X})\}
 \end{aligned}\end{gather}
where $q_{ (1-\alpha)}$ is the (1-$\alpha$) quantile of $\mathcal{Z}^{ (m)}=\max_{\boldsymbol{X} \in \mathcal{X}} \left| \frac{\boldsymbol{g}^{ (m)}(\boldsymbol{X})-\bar{\boldsymbol{g}}(\boldsymbol{X}) }{\hat{Std}[\bar{\boldsymbol{g}}(\boldsymbol{X})]} \right|$.

An advantage of this joint credible band formulation is that by inverting Equation \ref{equ_joint}, we can derive Simultaneous Band Score probabilities, $P_{SimBaS} (\boldsymbol{X})$ which tests if the tumor gradient effect is significantly different from 0 ($\boldsymbol{g}(\boldsymbol{X})=0 \space \forall \boldsymbol{X} \in \mathcal{X}$).
 \begin{gather*}
\begin{aligned}
 P_{\text{SimBaS}} (\mathbf{X}) &= \frac{1}{M} \sum_{m=1}^M\left( \left| \frac{\bar{\mathbf{g}}(\mathbf{X})}{\widehat{\text{Std}} \{ \bar{\mathbf{g}}(\mathbf{X}) \}} \right| \leq \mathcal{Z}^{(m)} \right) \\
 P_{\text{global}} &= \min_{\mathbf{X}} \{ P_{\text{SimBaS}} (\mathbf{X}) \}
\end{aligned}
\end{gather*}
\FloatBarrier
\begin{algorithm}[ht]
\caption{Procedure for estimating spatial decay $\phi$}\label{algo_phi}
\begin{algorithmic}[1]
  \State \textbf{Input:} $\boldsymbol{y},\,\boldsymbol{B},\,\boldsymbol{\mathcal{S}},\,M,\,\boldsymbol{\phi_{range}},\,\alpha_{\text{test}}$
  \State Compute OLS estimates $\boldsymbol{\hat{\beta}_{\text{ols}}}$ for $\boldsymbol{y \sim B}$ and residuals $\boldsymbol{y}^{resid}=\boldsymbol{y}-\boldsymbol{B}\boldsymbol{\hat{\beta}_{\text{ols}}}$
  \State Split data into training and testing sets with sizes $n_{\text{train}}=(1-\alpha_{\text{test}})\times N$ and $n_{\text{test}}=\alpha_{\text{test}}\times N$
  \For{$\tilde{\phi} \in \boldsymbol{\phi_{range}}$}
    \State Sample  $\boldsymbol{\tilde{\gamma}}=\{\tilde{\sigma}^2_y,\tilde{\tau}^2\}$ from a spatial-only robust adaptive Metropolis--Hastings algorithm with likelihood:
    \State $\boldsymbol{y^{resid}_{\text{train}}}\sim N\!\left(\boldsymbol{0},\,\tilde{\sigma}^2_y \boldsymbol{I}_{n_{\text{train}}}+\tilde{\tau}^2\boldsymbol{C_{\tilde{\phi}}(\mathcal{S}_{\text{train}})}\right)$
    \For{$m = 1, \ldots, M$ }
      \State $\boldsymbol{\Sigma}^{(m)}=\left(\tilde{\sigma}^{2(m)}_y*I_{N}+\tilde{\tau}^{2(m)}C_{\tilde{\phi}}(\mathcal{S})\right)$
      \State Calculate the posterior predictive mean of  $\boldsymbol{y}^{resid}_{\text{test}}$ given $\boldsymbol{y}^{resid}_{\text{train}}$:
      \Statex \hspace{\algorithmicindent}%
            $\boldsymbol{\mu}^{resid,(m)}_{\text{test}}
            =\boldsymbol{\Sigma}^{(m)}_{\text{test,train}}
            \big[\boldsymbol{\Sigma}^{(m)}_{\text{train,train}}\big]^{-1}
            \boldsymbol{y}^{resid}_{\text{train}}$
    \EndFor
    \State Average $\boldsymbol{{\mu}}^{resid,(m)}_{\text{test}}$ over MCMC draws:
    $\boldsymbol{\bar{\mu}}^{resid}=\frac{1}{M}\sum_{m=1}^M\boldsymbol{{\mu}}^{resid,(m)}_{\text{test}}$
    \State Calculate the root mean squared error:
    $RMSE_{\tilde{\phi}}=\sqrt{\frac{1}{n_{\text{test}}}\sum_{i=1}^{n_{\text{test}}}\left({{y}^{resid}_{\text{test},i}}-\bar{\mu}^{resid}_{i}\right)^2}$
  \EndFor
  \State \textbf{Output:} ${\phi_{\text{best}}} = \operatorname*{argmin}_{\tilde{\phi} \in \boldsymbol{\phi_{range}}} \left(RMSE_{\tilde{\phi}}\right)$
\end{algorithmic}
\end{algorithm}

\begin{figure}[ht]
    \centering
    \includegraphics[width=1\linewidth]{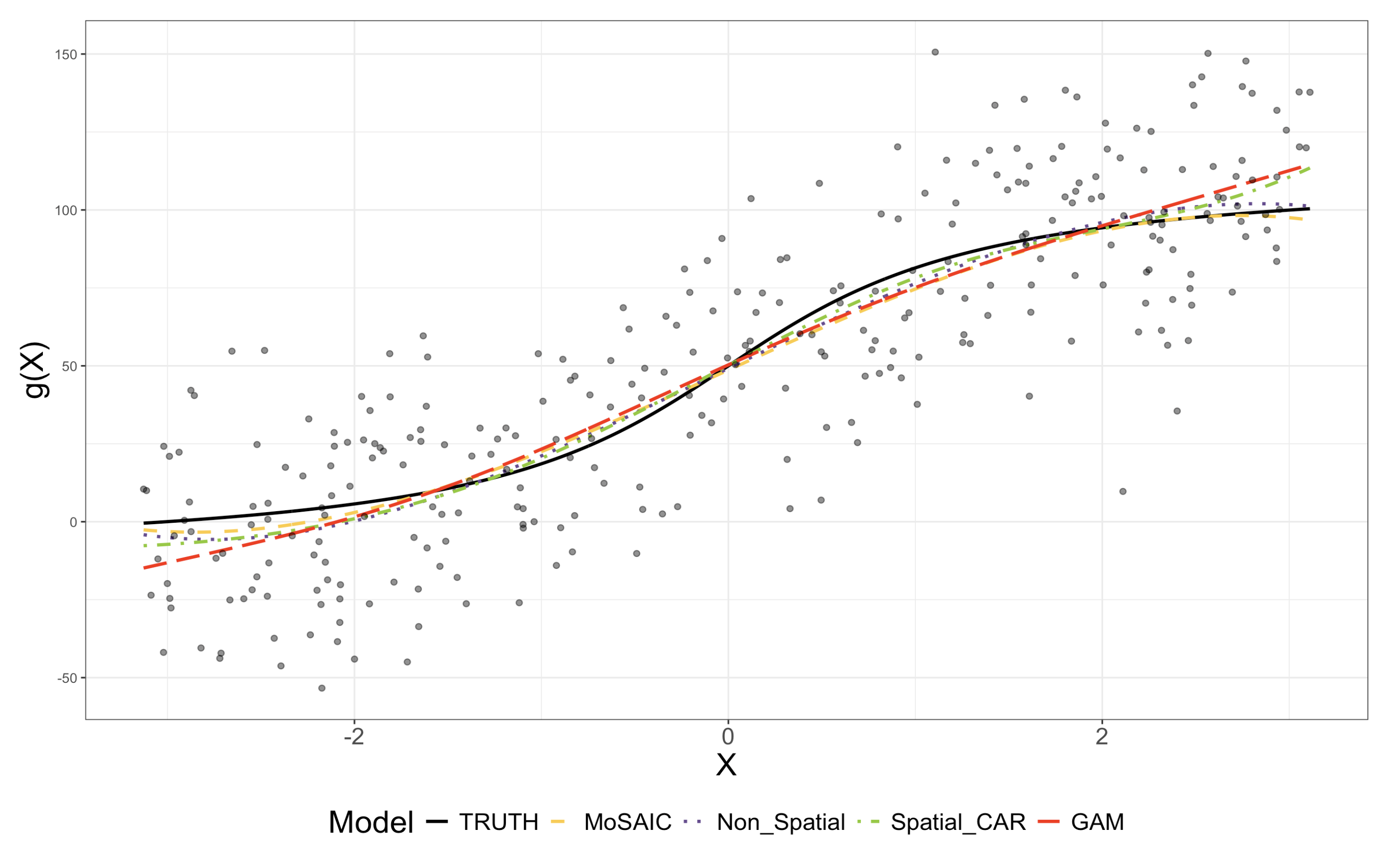}
    \caption{Plot of the true non-linear covariate effect (seed=5) and the estimated curve across all 4 models. The grey points are the observed data points and the black line represents the truth. While Non-Spatial has lower MSE then MoSAIC, their curves are very similar in shape}
    \label{supp:fittedcurve}
\end{figure}

\begin{algorithm}[ht]
\caption{Adaptive Metropolis Update}\label{algo_mcmc}
\begin{algorithmic}[1]

\Require $\boldsymbol{\gamma_{n-1}}$, $\boldsymbol{S_{n-1}}$, target acceptance $\alpha_{\text{target}} = 0.235$,$n_{adapt}$

\State Draw $\mathbf{U}_n \sim \mathcal{N}(0, I)$
\State Compute proposal:
      \[
      \mathbf{Z}_n \gets \boldsymbol{\gamma_{n-1}} + \boldsymbol{S_{n-1}} \mathbf{U}_n
      \]

\State Compute acceptance probability:\[
      \alpha_n \gets \min\left(1,\ \frac{\pi(\mathbf{Z}_n)}{\pi(\boldsymbol{\gamma_{n-1}})}\right)
      \]
\State $\boldsymbol{\gamma_n} \gets
\begin{cases}
    \mathbf{Z}_n & \text{With probability } \alpha_n\\
    \boldsymbol{\gamma_{n-1}} & \text{otherwise}\\
\end{cases}$
\State For  $n\le n_{adapt}$, update covariance factor $\boldsymbol{S_n}$ such that:
      \[
      \boldsymbol{S_n} \boldsymbol{S_n}^T = 
      \boldsymbol{S_{n-1}} \, (I + n^{-2/3}(\alpha_n - \alpha_{\text{target}})
          \frac{\mathbf{U}_n \mathbf{U}_n^T}{\|\mathbf{U}_n\|^2}) \, \boldsymbol{S_{n-1}}^T
      \]
\State \Return $\boldsymbol{\gamma_n}, \boldsymbol{S_n}$

\end{algorithmic}
\end{algorithm}
\FloatBarrier
\newpage
\suppsection{Simulations}\label{supp:sim}
\subsuppsection{Calculation of Predictive Coverage and Mean Squared Predictive Error}\label{supp.predictive.Coverage}
\FloatBarrier
In simulation settings, we assessed the predictive performance of our models via the 95\% predictive coverage and mean squared predictive error. First, we split the dataset into a training and testing set, withholding 30 observations for testing. We fit each model to the training data and used these MCMC samples to predict the outcomes of the test data $\boldsymbol{y^{test}}$. We make model based predictions of our testing data for the \texttt{MoSAIC} and Non-Spatial models. For the Spatial\_CAR model, we cannot make model based predictions of the spatial random effect at new locations. We use inverse distance weight interpolation for the CAR model in order to predict outcomes at new locations (\citealt{IDW}). The 95\% predictive coverage and mean squared predictive error use the resulting predictions, $\boldsymbol{y}^{*(m)}$, from Algorithm \ref{alg:unIfied}. The mean squared predictive error is defined as follows
\begin{gather*}
    \begin{aligned}
        MSPE&=\frac{1}{n_{test}\times M}\sum_{m=1}^{M}\sum_{k=1}^{n_{test}}({y^{test}_{k}}-{y^{*(m)}_k})^2.
    \end{aligned}
\end{gather*}
Predictive coverage is the probability that the observed outcomes are captured by the estimated credible bands. The nominal coverage is 95\%. Given $M$ predicted outcomes for each test observation,$\boldsymbol{y}^{*}_{i}=\{y_i^{*(1)},\dots,y_i^{*(M)}\}$, we calculate the $100(1-\alpha)$\% predictive coverage ($\bar{PC}_{1-\alpha}$) as follows. 
\begin{gather*}
    \begin{aligned}
         L_i,U_i \gets& \text{Quantiles}(\boldsymbol{y^{*}_{i}}; \{\alpha/2\}, \{ 1-\alpha/2\})\\
    \bar{PC}_{1-\alpha} &= \frac{1}{n_{test}}\sum_{i=1}^{n_{test}} (y^{\text{test}}_i \in [L_i, U_i]).
    \end{aligned}
\end{gather*}
\subsuppsection{Fitted covariate curves in the non-linear simulations}\label{supp.sim.curves}
Figure \ref{supp:fittedcurve} shows the fitted covariate curves for one of the simulation data sets in the non-linear scenario (Scenario 3). The gray dots are the observed data and the black line represents the true $g(\boldsymbol{X})$. Non-spatial model has slightly lower MSE than \texttt{MoSAIC} but visually the curves are very similar. 

\subsuppsection{Recovering $\boldsymbol{\beta}$}\label{supp.betas}
Using samples of $\{\sigma_x^2,\sigma_y^2,\tau^2,\sigma_Z^2\}$, we jointly sample any patient-specific estimates ($\boldsymbol{ \mu}$), biomarker estimates ($\boldsymbol{\theta}$), and latent spatial effects ($\boldsymbol{\psi}$) from their posteriors. Thus, $\boldsymbol{\beta^{T}}=[
 \boldsymbol{\mu^T},
 \boldsymbol{\theta^T},
 \boldsymbol{\psi}^T]$ is sampled from
\begin{gather*}\begin{aligned}
 &p (\boldsymbol{\beta|y,Z,B}, \tau^2, {\sigma^2_x},\sigma^2_y,\sigma^2_Z) \sim N (\boldsymbol{V}
 \begin{bmatrix}
 \boldsymbol{Z^T\Sigma_y^{-1}y}\\
 \boldsymbol{ B^T\Sigma_y^{-1}y}\\
  \boldsymbol{\Sigma_y^{-1}y}
 \end{bmatrix},
 \boldsymbol{V})\\
&\boldsymbol{V} = 
\operatorname{blockdiag}\!\left[
\left(\boldsymbol{Z}^T \boldsymbol{\Sigma}_y^{-1} \boldsymbol{Z} 
      + \frac{1}{\sigma_Z^2}\boldsymbol{I_I}\right)^{-1},
\;
\left(\boldsymbol{B}^T \boldsymbol{\Sigma}_y^{-1} \boldsymbol{B}
      + {\sigma_x^{-2}}\boldsymbol{K^{-1}}\right)^{-1},
\;
\left(\boldsymbol{\Sigma}_y^{-1} 
      + \frac{1}{\tau^2} \boldsymbol{C}_\phi^{-1}\right)^{-1}
\right].
 \end{aligned}
 \label{shrink_eq}
 \end{gather*} 
\subsuppsection{Summary Table for Simulation Results}\label{supp.sim.table}
The simulation results across three distinct data scenarios—Non-Linear, High Spatial Dependence, and Low Spatial Dependence—demonstrate the comparative performance of the \texttt{MoSAIC} model against Spatial (CAR), Non-Spatial, and GAM frameworks. Performance was evaluated using Watanabe-Akaike Information Criterion (WAIC), Mean Squared Error (MSE), Mean Squared Prediction Error (MSPE), 95\% prediction coverage, and computation time. Table \ref{sim_table} includes the mean and standard deviations of WAIC, MSE, and MSPE under each simulation scenario. \texttt{MoSAIC} and Non-Spatial were the slowest overall but all models were fit in under 10 minutes when run on a laptop with M1 processor and 16GB of memory. R markdown tutorials for each simulation setting can be found in the Simulations folder\anonflag{ at \href{https://github.com/jcaldous/MoSAIC}{jcaldous/MoSAIC}} on Github. 
\begin{table}[t]
\caption{Simulation Results Table}\label{sim_table}
\fontsize{12.0pt}{14.0pt}\selectfont
\begin{tabular*}{\linewidth}{@{\extracolsep{\fill}}lccccc}
\toprule
\textbf{Characteristic} & \textbf{WAIC} & \textbf{MSE} & \textbf{MSPE}& \makecell{\textbf{Prediction}\\ \textbf{Coverage}(95\%)}& Time to Fit\\
\midrule\addlinespace[2.5pt]
\multicolumn{6}{l}{Non-Linear} \\[2.5pt] 
\midrule\addlinespace[2.5pt]
\texttt{MoSAIC} & \textbf{2020 (47.9)}& 36.2(21.4)&\textbf{342(74.9)} & \textbf{0.95 (0.04)} & 6 min\\ 
Spatial (CAR) & 2456 (73.8) & 38.2 (25.7) & 1216(295) & 0.91 (0.07) & 8 sec\\ 
Non-Spatial & 2400 (41.7) & \textbf{33.3 (20.9)} & 864 (17) & 0.94 (0.05) & 4 min \\ 
GAM & - & 77.4 (108) & 629 (188) & \textbf{0.95 (0.04)} & <1 sec\\ 
\midrule\addlinespace[2.5pt]
\multicolumn{6}{l}{High Spatial Dependence} \\[2.5pt] 
\midrule\addlinespace[2.5pt]
\texttt{MoSAIC} & \textbf{1921 (33.3)} & \textbf{1.40 (1.51)} & \textbf{209 (50.6)} & \textbf{0.95 (0.05)}&6 min \\ 
Spatial (CAR) & 2379 (107) & 3.60 (3.35) & 1090 (276) & 0.86 (0.11)&  8 sec\\ 
Non-Spatial & 2300 (47.2) & 3.01 (2.74) & 599 (120) & \textbf{0.95 (0.04)}&4 min  \\ 
GAM & - & 2.28 (3.45) & 577 (192) & \textbf{0.95 (0.04)}& <1 sec\\ 
\midrule\addlinespace[2.5pt]
\multicolumn{6}{l}{Low Spatial Dependence} \\[2.5pt] 
\midrule\addlinespace[2.5pt]
\texttt{MoSAIC} & \textbf{1940 (31.7)} & 1.33 (1.27) & \textbf{163 (22.4)} & \textbf{0.95 (0.04)}&6 min \\ 
Spatial (CAR) & 2045 (198) & 1.03 (0.93) & 335 (90.4) & 0.87 (0.10)& 8 sec\\ 
Non-Spatial & 2002 (27.2) & 1.46 (1.39) & 193 (28.8) & \textbf{0.95 (0.04)}& 4 min \\ 
GAM & - & \textbf{0.68 (0.96)} & 181 (55.1) & \textbf{0.95 (0.04)}& <1 sec\\ 
\bottomrule
\end{tabular*}
\begin{minipage}{\linewidth}\textsuperscript{\textit{1}}Mean (SD)\\
\end{minipage}
\end{table}
\begin{algorithm}[t]
\caption{Predicting Outcomes $\boldsymbol{Y^{*}}$ for (\texttt{MoSAIC} and Spatial\_CAR)}
\label{alg:unIfied}
\footnotesize 
\begin{algorithmic}[1] 
\State Model $\in \{\text{MoSAIC}, \text{Spatial\_CAR}\}$, $\boldsymbol{X}_{\text{test}}$, $\boldsymbol{y}^{\text{test}}$, $M, \mathcal{S}$, $\boldsymbol{\mathbf{ID}}=[\boldsymbol{\mathbf{ID}^{train}},\boldsymbol{\mathbf{ID}^{test}}], \boldsymbol{\gamma}_{model}$. 
\For{$m = 1\dots M$} 
  \If{\texttt{MoSAIC}} ${\gamma}^{(m)}_{model}=(\boldsymbol{\theta^{(m)}}, rI^{(m)}_{\text{train}}, w^{(m)}, \sigma_y^{2(m)}, \tau^{2(m)}, \sigma_{RI}^{2(m)})$
    \State $\boldsymbol{C_\phi(\mathcal{S)}}=\begin{bmatrix}
      \boldsymbol{C_{\text{train}}}&\boldsymbol{C^T_{\text{test,train}}}\\
      \boldsymbol{C_{\text{test,train}}}&\boldsymbol{C_{\text{test}}}
    \end{bmatrix}$.
    \State $\boldsymbol{\mu}_{\text{fixed}}^{(m)} \gets \boldsymbol{X_{\text{test}}}\boldsymbol{\theta^{(m)}}$
    \For{$i = 1, \dots, n_{\text{test}}$}
      \If{$\mathrm{ID}^{\text{test}}_i \in \mathbf{ID^{\text{train}}}$}
        \State $rI^{(m)}_i \gets rI^{(m)}_{\text{train}, j}$ with $j$ such that $\mathrm{ID}^{\text{test}}_i = \mathrm{ID}^{\text{train}}_j$
      \Else
        \State $rI^{(m)}_i \gets \mathcal{N}\!\left(0, \left(\sigma_{RI}^{(m)}\right)^2\right)$
      \EndIf
    \EndFor
    \State $\boldsymbol{w^{*(m)}} \gets \mathcal{N}\!\Big(\boldsymbol{C_{\text{test,train}} C_{\text{train}}^{-1} w^{(m)}},
      \tau^{2(m)} \boldsymbol{( C_{\text{test}} - C_{\text{test,train}} C_{\text{train}}^{-1} C_{\text{test,train}}^T)}\Big)$ 
    \State $\boldsymbol{\mu}_{\text{pred}}^{(m)} \gets \boldsymbol{\mu}_{\text{fixed}}^{(m)} + \boldsymbol{rI}^{(m)} + \boldsymbol{w}^{*(m)}$; $V_{\text{pred}}^{(m)} \gets \sigma_y^{2(m)}$
  \EndIf

  \If{Spatial\_CAR} ${\gamma}^{(m)}_{model}=(\boldsymbol{\theta^{(m)}}, \phi^{(m)}_{\text{train}}, \sigma^{2(m)}_y, \tau^{2(m)})$, Distance Matrix $\boldsymbol{D}$.
    \State $\boldsymbol{\mu}_{\text{fixed}}^{(m)} \gets \boldsymbol{X_{\text{test}}}\boldsymbol{\theta^{(m)}}$
    \For{$i = 1, \dots, n_{\text{test}}$}
      \State $w_{ij} \gets (D_{ij} + \epsilon)^{-2}$; $\phi^{*(m)}_i \gets \frac{\sum_{j} w_{ij} \phi^{(m)}_j}{\sum_{j} w_{ij}}$
    \EndFor
    \State $\boldsymbol{\mu}_{\text{pred}}^{(m)} \gets \boldsymbol{X_{\text{test}}}\boldsymbol{\theta^{(m)}} + \boldsymbol{\phi}^{*(m)}$; $V_{\text{pred}}^{(m)} \gets \sigma^{2(m)}_y + \tau^{2(m)}$ 
  \EndIf

  \State $\boldsymbol{y^{*(m)}} \gets \mathcal{N}(\boldsymbol{\mu}_{\text{pred}}^{(m)}, V_{\text{pred}}^{(m)} I_{n_{\text{test}}})$
\EndFor
\State \textbf{Output Predictions:} $\boldsymbol{Y^{*}}=\{\boldsymbol{y^{*(1)}},\dots, \boldsymbol{y^{*(m)}},\dots, \boldsymbol{y^{*(M)}}\}$
\end{algorithmic}
\end{algorithm}

\FloatBarrier
\newpage
\suppsection{Sarcomatoid Renal Cell Carcinoma Analysis}\label{supp:RDA}
\subsuppsection{Tumor gradient and Immune Cell Definitions for Real Data Analysis}\label{supp.PDL1.summ}
\FloatBarrier
The expression of PD-L1 and NCAD are measured across pixels of the cell and summarized into 5 distinct values: minimum, maximum, mean, total, and standard deviation. NCAD expression at the FOV level was defined as the average of mean cellular NCAD expression. Since PD-L1 expression is generally very low as seen in Figure \ref{fig:PDL1}, we defined FOV level PD-L1 expression as the sum of the total cellular PD-L1 across all cells within an FOV. The FOV level PD-L1 expression is rescaled to similar to NCAD expression by dividing by $1000$. Table \ref{supp.celltable} outlines the immunophenotyping criteria for cells. These cellular phenotypes were used to define the pairwise cell colocalizations.
\begin{table}[ht]
\centering
\caption{Cell type definitions.}
\label{supp.celltable}
\begin{tabular}{@{}ll@{}}
\toprule
\textbf{Cell type} & \textbf{Markers} \\
\midrule
Helper T     & CD3+ and CD8- and CD163- \\
Cytotoxic T  & CD8+ and CD163- \\
Macrophage   & CD163+ \\
Tumor        & All remaining cells \\
\bottomrule
\end{tabular}
\end{table}
\FloatBarrier
\subsuppsection{Quantifying cell clustering within an FOV}\label{supp.clustering}
 For our real data analysis, we used DIMPLE software to measure each pairwise cellular co-localization(\citealt{MASOTTI2023100879}). DIMPLE first generates a cell intensity function for each cell type and then calculates the pairwise distance between two cell intensities. The Jensen-Shannon distance is an ideal choice for our setting because it is scientifically easy to interpret and is not sensitive to image artifacts (i.e. holes or tears). The Jensen-Shannon distance makes modeling cell co-occurrence more interpretable since we know a value of 0 indicates perfect overlap and 1 complete separation. We rescale the distance such that for a given dimple distance $d_{dimple}$, 
 \begin{align*}
    {y_{dimple}}&=100*(1-d_{dimple}).
 \end{align*}
 Additionally, DIMPLE software requires no permutations and produces output ready for downstream modeling, making it efficient and well suited for our modeling needs. DIMPLE requires users to specify the grid size and smoothing distance in order to generate the initial cell-type intensities. In all our modeling, the grid is set to roughly the size of a cell (5 to 10 $\mu m$) and the smoothing is set to be between 20-60 $\mu m$, which is considered the sphere of influence of a cell (\citealt{MASOTTI2023100879}).

\begin{figure}
    \centering
    \includegraphics[width=1\linewidth]{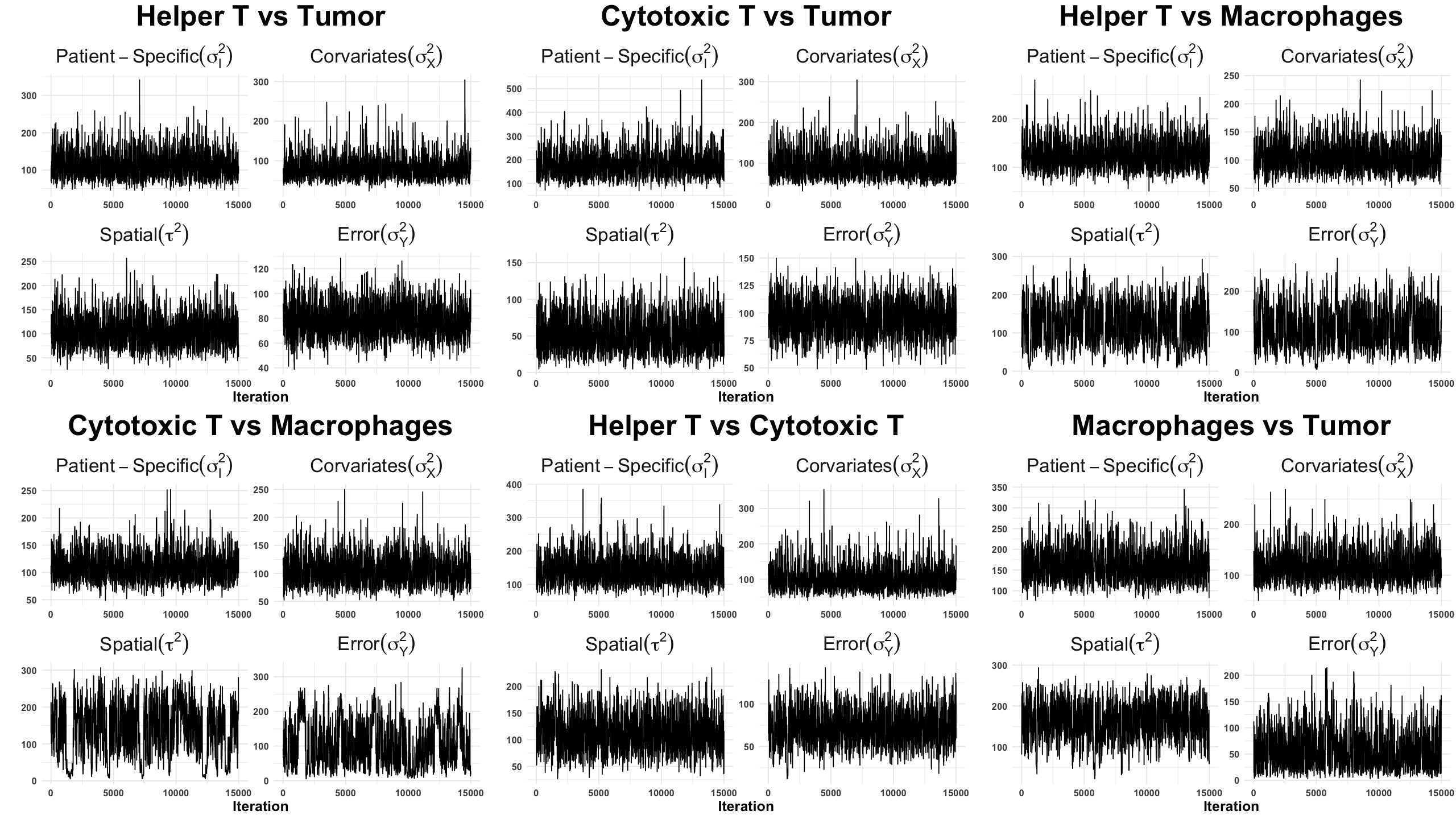}
    \caption{Traceplots for pairwise colocalization models}
    \label{fig:trace}
\end{figure}

\subsuppsection{\texttt{MoSAIC} MCMC Convergence and Model Diagnostics}\label{supp.diagnostics}
Convergence of our models is assessed visually through the MCMC plots of the variance terms, which show our model is exploring the posterior well (Figure \ref{fig:trace}). Table \ref{supp:summary} includes additional information about the MCMC samples of the variance terms. For each model, we report the posterior means and point-wise confidence intervals, the effective sample size(n effective) out of $15,000$ and the Geweke Score. Absolute Geweke scores below $1.96$ indicate no strong evidence against convergence. An n effective lower than $15,000$ indicates larger autocorrelation between our samples. The n effective and Geweke scores are derived via the respective function in the coda R package (\citealt{coda}).

Table \ref{supp:modelfit} compares the model fit of \texttt{MoSAIC} and the non spatial alternative via DIC and WAIC. The lower scores across all 6 models show the spatial \texttt{MoSAIC} model consistently fit the data better than when the spatial effect was omitted. 
\FloatBarrier
\begin{table}[t]
\caption{Posterior Means and Credible intervals for \texttt{MoSAIC} models applied to RCC data}\label{supp:summary}
\centering
\footnotesize
\setlength{\tabcolsep}{3pt}
\renewcommand{\arraystretch}{0.92}
\begin{tabular}{@{}cc|cccc@{}}
\toprule
\textbf{Model} & \textbf{Metric} & $\sigma^2_Y$ & $\tau^2$ & $\sigma^2_I$ & $\sigma^2_x$ \\
\midrule

\multirow{3}{*}{\makecell[c]{Helper T vs \\ Tumor}}
 & \makecell{Posterior Mean\\ (CI)} & \makecell{77.7 \\ (55.9, 102.8)} & \makecell{103.1 \\ (54.2, 169.1)} & \makecell{110.8 \\ (63.3, 187.5)} & \makecell{78.7 \\ (43.7, 142.4)} \\
 & N effective & 1042.6 & 1021.8 & 1066.2 & 974.1 \\
 & Geweke & 0.40 & -0.20 & 1.80 & -1.80 \\
\midrule

\multirow{3}{*}{\makecell[c]{Helper T vs \\ Macrophage}}
 & \makecell{Posterior Mean\\ (CI)} & \makecell{107.0 \\ (27.1, 213.4)} & \makecell{119.3 \\ (19.4, 219.7)} & \makecell{124.0 \\ (82.2, 186.6)} & \makecell{104.0 \\ (66.8, 158.9)} \\
 & N effective & 393.6 & 339.5 & 1023.7 & 1021.8 \\
 & Geweke & 1.40 & -0.80 & 1.10 & 0.00 \\
\midrule

\multirow{3}{*}{\makecell[c]{Helper T vs \\ Cytotoxic}}
 & \makecell{Posterior Mean\\ (CI)} & \makecell{71.4 \\ (35.0, 111.2)} & \makecell{110.3 \\ (53.3, 181.1)} & \makecell{138.7 \\ (79.9, 223.7)} & \makecell{101.8 \\ (57.4, 186.2)} \\
 & N effective & 921.4 & 919.7 & 1034.5 & 876.8 \\
 & Geweke & -1.00 & 1.90 & -1.30 & 0.20 \\
\midrule

\multirow{3}{*}{\makecell[c]{Cytotoxic vs \\ Macrophage}}
 & \makecell{Posterior Mean\\ (CI)} & \makecell{104.1 \\ (15.2, 232.9)} & \makecell{133.5 \\ (13.4, 248.7)} & \makecell{108.8 \\ (71.4, 163.5)} & \makecell{104.2 \\ (67.1, 159.1)} \\
 & N effective & 163.2 & 157.7 & 899.7 & 1042.2 \\
 & Geweke & 0.80 & -0.70 & 0.20 & 0.10 \\
\midrule

\multirow{3}{*}{\makecell[c]{Cytotoxic vs \\ Tumor}}
 & \makecell{Posterior Mean\\ (CI)} & \makecell{93.6 \\ (65.8, 124.1)} & \makecell{50.5 \\ (13.8, 98.6)} & \makecell{176.9 \\ (102.5, 289.1)} & \makecell{93.2 \\ (51.9, 165.1)} \\
 & N effective & 991.8 & 1038.9 & 992.2 & 1043.9 \\
 & Geweke & 0.30 & 0.10 & -1.20 & 0.80 \\
\midrule

\multirow{3}{*}{\makecell[c]{Macrophage vs \\ Tumor}}
 & \makecell{Posterior Mean\\ (CI)} & \makecell{55.1 \\ (8.5, 145.3)} & \makecell{159.3 \\ (60.9, 233.5)} & \makecell{160.1 \\ (101.4, 243.2)} & \makecell{116.3 \\ (74.1, 182.1)} \\
 & N effective & 479.3 & 444.4 & 962.5 & 1059.8 \\
 & Geweke & 0.60 & -0.00 & -0.50 & -1.30 \\
\bottomrule
\end{tabular}
\end{table}
\begin{table}[t]
\caption{Model fit metrics for \texttt{MoSAIC} vs Non-Spatial alternative in real data setting}\label{supp:modelfit}
\centering
\small
\renewcommand{\arraystretch}{0.92}
\setlength{\tabcolsep}{4pt}
\begin{tabular}{ccll}
  \hline
  Pairwise Cellular Colocalization & Model & DIC & WAIC \\
  \hline
  \multirow{2}{*}{Helper T vs Tumor} & \texttt{MoSAIC} & \textbf{2454} & \textbf{2459} \\
   & Non-Spatial & 2544 & 2543 \\
  \hline
  \multirow{2}{*}{Cytotoxic T vs Tumor} & \texttt{MoSAIC} & \textbf{2463} & \textbf{2472} \\
   & Non-Spatial & 2498 & 2500 \\
  \hline
  \multirow{2}{*}{Helper T vs Macrophages} & \texttt{MoSAIC} & \textbf{1681} & \textbf{1710} \\
   & Non-Spatial & 1787 & 1788 \\
  \hline
  \multirow{2}{*}{Cytotoxic T vs Macrophages} & \texttt{MoSAIC} & \textbf{1671} & \textbf{1738} \\
   & Non-Spatial & 1838 & 1839 \\
  \hline
  \multirow{2}{*}{Helper T vs Cytotoxic T} & \texttt{MoSAIC} & \textbf{2311} & \textbf{2323} \\
   & Non-Spatial & 2433 & 2435 \\
  \hline
  \multirow{2}{*}{Macrophages vs Tumor} & \texttt{MoSAIC} & \textbf{1581} & \textbf{1633} \\
   & Non-Spatial & 1835 & 1835 \\
  \hline
\end{tabular}
\end{table}
\FloatBarrier

\end{document}